\documentclass[useAMS]{mn2e}
\usepackage{psfig,epsfig}
\usepackage{graphicx}
\usepackage{amssymb}
\usepackage{epstopdf}

\begin{document}
\def\simlt{\mathrel{\rlap{\lower 3pt\hbox{$\sim$}}
        \raise 2.0pt\hbox{$<$}}}
\def\simgt{\mathrel{\rlap{\lower 3pt\hbox{$\sim$}}
        \raise 2.0pt\hbox{$>$}}}

\title[The role of galaxy mass on AGN emission]{The role of galaxy mass on AGN emission: a view from the VANDELS survey}

\author[Manuela Magliocchetti et al.]
{\parbox[t]\textwidth{M. Magliocchetti$^{1}$, L. Pentericci$^{2}$, M. Cirasuolo$^{3}$, G. Zamorani$^{4}$, R. Amorin$^{5,6}$, A. Bongiorno$^{2}$, A. Cimatti$^{7}$, A. Fontana$^{2}$, B. Garilli$^{8}$, A. Gargiulo$^{8}$, N.P. Hathi$^{9}$, D.J. McLeod$^{10}$, R.J. McLure$^{10}$,  M. Brusa$^{7}$,  A. Saxena$^{2}$, M. Talia$^{7}$}\\
\\
{\tt $^1$ INAF-IAPS, Via Fosso del Cavaliere 100, 00133 Roma,
  Italy}\\
  {\tt $^2$ INAF-OAR, Via Frascati 33, Monte Porzio Catone (RM), Italy }\\
 {\tt $^3$ European Southern Observatory, Karl-Schwarzschild-Str. 2, D-85748 Garching, Germany }\\
 {\tt $^4$    INAF-Osservatorio di Astrofisica e Scienza dello Spazio di Bologna, via Gobetti 93/3, 40129, Bologna, Italy}\\
  {\tt $^5$ Instituto de Investigaci—n Multidisciplinar en Ciencia y Tecnolog'a,
Universidad de La Serena, Raul Bitr‡n 1305, La Serena, Chile}\\
{\tt $^6$ Departamento de F'sica y Astronom'a, Universidad de La Serena, Av. Juan
Cisternas 1200 Norte, La Serena, Chile}\\
  {\tt $^7$ Department of Physics and Astronomy (DIFA), University of Bologna, Via Gobetti 93/2, I-40129 Bologna, Italy}\\    
      {\tt $^8$ INAF-IASF, via Bassini 15, 20133, Milano, Italy}\\
   {\tt $^9$ Space Telescope Science Institute, 3700 San Martin Drive, Baltimore, MD 21218, USA}\\
  {\tt $^{10}$ Institute for Astronomy, University of Edinburgh, Royal Observatory, Edinburgh EH9 3HJ, UK}\\
 }
 \maketitle
 \begin{abstract}
 We present a comparative analysis of the properties of AGN emitting at radio and X-ray wavelengths. 
 The study is performed on 907 X-ray AGN and 100 radio AGN selected on the CDFS and UDS fields and makes use of new and ancillary data available to the VANDELS collaboration.
 Our results indicate that the mass of the host galaxy is a fundamental quantity which determines the level of AGN activity at the various wavelengths. 
Indeed large stellar masses are found to be connected with AGN radio emission, as virtually all radio-active AGN reside within galaxies of  $M_*> 10^{10} M_\odot$. Large stellar masses also seem to favour AGN activity in the X-ray, even though X-ray AGN present a mass distribution which is  more spread out and with a non-negligible tail at $M_*\simlt 10^{9} M_\odot$.
 Stellar mass alone is also observed to play a fundamental role in simultaneous radio and X-ray emission: the percentage of AGN active at both wavelengths increases from around 1\% of all X-ray AGN residing within hosts of $M_*<10^{11} M_\odot$ to $\sim 13$\% in more massive galaxies. 
In the case of radio-selected AGN, such a percentage moves from $\sim 15$\%  to $\sim 45$\% (but up to $\sim 80$\% in the deepest fields). 
Neither cosmic epoch, nor radio luminosity, X-ray luminosity, Eddington ratio or star-formation rate of the hosts are found to be connected to an enhanced probability for joint radio+X-ray emission of AGN origin. Furthermore, only a loose relation is observed between X-ray and radio luminosity in those AGN which are simultaneously active at both frequencies.

  \end{abstract}


 \section{Introduction}
 AGN come in all kinds of flavour and emit in all bands of the electromagnetic spectrum. However, the overlapping between different classes of AGN detected at the different wavelengths is small. Sometimes very small, especially when it comes to sources active at radio frequencies  (e.g. Hickox et al. 2009).  

Apart from the obvious issue of  depth of the observations, the detectability of an AGN within a particular band is the result of a number of complex factors. These range from the physics of the AGN itself (i.e. radiatively efficient mode vs radiatively inefficient or jetted "radio" mode, e.g. Merloni, Heinz \& Di Matteo 2003; Croton et al. 2006; Heckman \&  Best 2014), to the lifetime of the AGN phenomenon which seems to be  at least a factor $\sim 10$ longer in the radiatively efficient phase with respect to the radiatively inefficient one (e.g. Porciani, Magliocchetti \& Norberg 2004;  Blundell \& Rawlings 1999;  King \& Nixon 2015; Schawinski et al. 2015; Schmidt et al. 2017; Magliocchetti et al. 2017), although very recent results by Heesen et al. (2018) based on LOFAR observations seem to provide estimates for the lifetime of the radio phase which are roughly a factor  $\sim 10-20$ higher than what is currently believed. Amongst other factors there are the mass of the black hole, as radio-active AGN are mostly found to be powered by black holes with $M_{BH}\simgt 10^9 M_\odot$ while radio-quiet ones by black holes of lower masses (e.g. Jarvis \& Mc Lure 2002; Metcalf \& Magliochetti 2006),  the orientation of the AGN and of the surrounding torus with respect to the observer (e.g.  Antonucci 1993) and the amount and status of the dust in which the AGN is embedded  (see Hickox \& Alexander 2018 for an extensive review on obscured AGN). 

In addition, host galaxy properties and environment also seem to be tightly related to AGN activity. It is in fact nowadays clear that AGN emitting in the radio band are mainly associated not only to the most massive galaxies (e.g. Heckman \& Best 2014; Magliocchetti et al. 2018; Sabater et al. 2019) but also to the most massive dark matter halos, of sizes comparable only to those which host groups-to-clusters of galaxies (e.g. Magliocchetti et al. 2004; 2017; 2018b; Hatch et al. 2014; Retana-Montenegro et al. 2017). The same also seems to be true for X-ray emitting AGN, although in this latter case there is a less general consensus, with values for the typical dark matter halos possibly dependent on obscuration (e.g. Hickox et al. 2009; DiPompeo et al. 2017; Powell et al. 2018). On the other hand, optical and infrared-bright AGN are largely found in relatively smaller galaxies and haloes (e.g. Porciani et al. 2004; Shen et al. 2009; Hickox et al. 2009; 2011; Retana-Montenegro et al. 2017). There are claims that the aforementioned large-scale environmental differences could be entirely attributed to host galaxy properties such as stellar mass and/or star-formation rate (e.g. Yang et al. 2018; Georgakakis et al. 2019). However, although it is nowadays clear that galaxy mass plays a crucial role in the interpretation of the various  results, this might not be the full story (e.g. Mendez et al. 2016; Magliocchetti et al. 2018b).

Within this complex scenario, particularly interesting are those sources which present simultaneous emission at different wavelengths as from their investigation it is possible to shed better light on the different processes and properties which characterize AGN emission. However, in order to proceed in a statistically meaningful way, it is mandatory to start from parent AGN samples which are as complete as possible. 




This is why in this work we will use a combination of deep and ultra-deep radio and X-ray observations coming from surveys performed on the Chandra Deep Field South (CDFS) and the UKIDSS Ultra Deep Field (UDS). These have been implemented by using new and ancillary data available to the VANDELS collaboration (McLure et al. 2018: Pentericci et al. 2018) in order to gather information on properties such as stellar mass and star-formation rate (SFR) of the galaxies hosts of radio and X-ray emitters. 
 
Such an approach is not new and many works can be found in the recent literature on cross identifications and studies of the properties of the galaxies host of AGN selected in the various bands of the electromagnetic spectrum (e.g. Brusa et al. 2009; Bongiorno et al. 2012; 2016; Aird et al. 2012; Georgakakis et al. 2014; 2017; Magliocchetti et al. 2014; 2016; 2018; Bonzini et al. 2015; Yang et al. 2017; Smolcic et al. 2017 just to mention a few). However, this is the first time that information on deep AGN surveys obtained on more than one field is combined with a homogeneous and complete sample of sources which provides reliable mass estimates.
 
 The layout of the paper is as follows: \S 2 presents the VANDELS survey, while \S 3 presents the radio and X-ray surveys  on the CDFS and UDS. 
 \S 4 describes the methods adopted to identify AGN both at X-ray and radio frequencies. The resulting samples are analyzed and discussed in \S 5 and \S 6. 
 \S 7 summarizes our conclusions.
 Throughout the paper we will adopt  a $\Lambda$CDM cosmology with $\Omega_0=0.3$, $\Lambda=0.7$ and $H_0=70$ km sec$^{-1}$ Mpc$^{-1}$. All magnitudes are AB magnitudes.

 \section{VANDELS}
 The VANDELS survey is a deep optical spectroscopic survey in the CANDELS CDFS and UDS 
 fields with the VIMOS spectrograph on the ESO's Very Large Telescope (VLT). 
 It targets  massive passive galaxies at $1.0\le z\le 2.5$, bright star-forming galaxies
at $2.4\le z\le 5.5$ and fainter star-forming galaxies at $3.0\le z\le 7.0$.
All galaxies are  drawn from four independent 
catalogues, called MASTER catalogues. The CDFS and UDS regions are covered by the CANDELS survey (Grogin et al. 2011; Koekemoer et al. 2011), and 
benefit from extensive WFC3/IR imaging: these areas are called CDFS-HST and UDS-HST. In these regions, the official photometric catalogues produced by the CANDELS team were used  as master catalogues (Galametz et al. 2013; Guo et al. 2013).  
The only differences are that the Ks\_HAWKI photometry has been updated to include the final HUGS survey data (see Fontana et al. 2014) and the photometry in the CTIO\_U band filter has been removed. Fluxes are corrected to total based on the ratio of F160W\_flux\_ISO and 
F160W\_flux\_AUTO (see Guo et al. 2013 for more details). The catalogue has been cut at $H_{AB}$ = 27.05.

Within the wider regions surrounding the CANDELS fields, new  PSF-homogenized catalogues were produced, primarily from publicly available ground-based imaging: these regions are called CDFS-GROUND and UDS-GROUND. For these catalogues  the fundamental photometry is based on 2$^{\prime\prime}$-diameter apertures measured on the PSF-homogenized images.
For more details on the available photometry with a list of all filters employed we refer the reader to  McLure et al. (2018).

Photometric redshifts for the CANDELS catalogues are available in Galametz et al. (2013), and Guo et al. (2013). For the ground-based targets, photometric redshifts based on the median of seven independent photometric redshift runs on the updated ground-based photometry catalogues were derived by the VANDELS team. The combined catalogue of derived parameters features stellar masses, dust attenuation, star-formation rates and rest-frame magnitudes (in all 32 broad-band filters used within VANDELS) obtained via SED template fits at the adopted photometric redshift or at the spectroscopic redshift whenever this is known.  The templates adopted for the SED fitting are the Bruzual \& Charlot (2003) stellar population models with solar metallicity, a Chabrier IMF and 
declining tau-model star-formation histories with values of tau in the range $0.3 < \tau < 10.0$ Gyr. The ages of the SED templates are 
allowed to vary within the range 50 Myr $<$ age $<$ Age of the Universe at the chosen redshift. Dust reddening is applied using the Calzetti et al. 
(2000) starburst attenuation curve, with $A_V$ values in the range $0.0 <A_V< 3.0$. The absorption due to the IGM is calculated via the 
Madau (1995) prescription. This range of templates  was chosen for its ability to produce sensible total star-formation rates (see e.g. Wuyts et al. 2011) and 
stellar masses (see McLure et al. 2018 for more details). 

The VANDELS survey was completed in February 2018.  Fully reduced  spectra
and ancillary products from the survey are made available  to the community
in annual data releases through the ESO website and our public webpage (http://vandels.inaf.it). The most recent release (DR2) was made publicly available in October 2018
and contains spectra for 1362 galaxies.
 \section{The Fields}
 
 \begin{figure}
\includegraphics[scale=0.4]{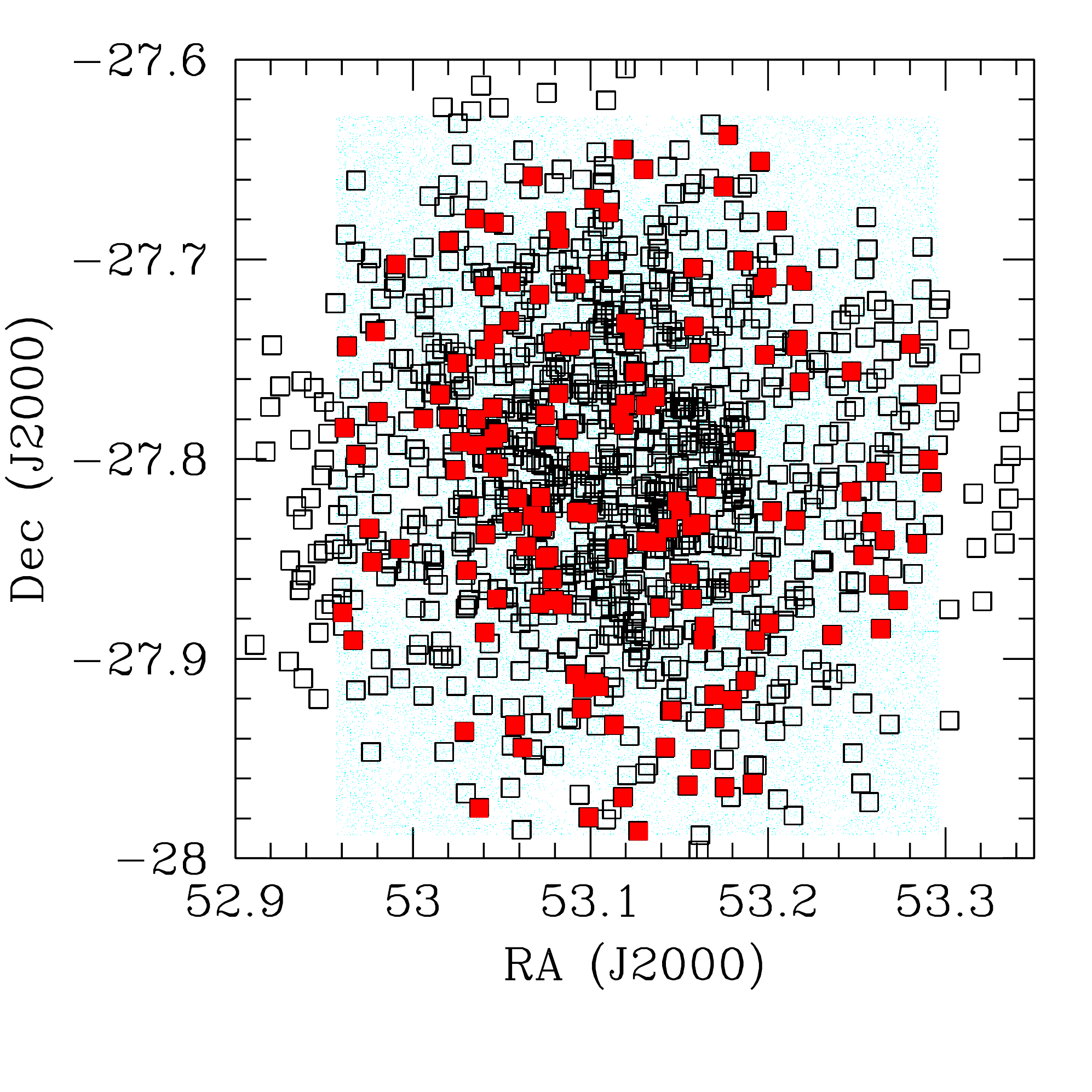}
\caption{Projected distribution onto the sky of X-ray detected sources from the CDFS (crosses). Filled squares indicate those objects with a 1.4 GHz counterpart. The shaded cyan  area shows the photometric coverage of the VANDELS MASTER CDFS catalogues (cft text for details).}
\end{figure}	

 \subsection{Chandra Deep Field South (CDFS)}
  
\begin{table*}
\begin{center}
\caption{Properties of the VANDELS-CDFS radio selected sample. AGN are classified  according to equation (1).}
\begin{tabular}{llllll}
 \hline
 & on area& with id & with zspec & with X-ray counterpart & with X-AGN counterpart\\
\hline
All & 355& 277& 186& 159& 115\\
AGN & 43 & 43& 21 &22& 17\\

\end{tabular}
\end{center}
\end{table*}

  
\begin{table*}
\begin{center}
\caption{Properties of the VANDELS-CDFS X-ray selected sample. AGN classification follows that of Luo et al. (2017).}
\begin{tabular}{llllll}
 \hline
 & on area& with id & with zspec& with radio counterpart & with radio-AGN counterpart\\
\hline
All & 996& 839& 650& 159&22\\
AGN & 680 & 562& 389&115&17\\

\end{tabular}
\end{center}
\end{table*}


\begin{table*}
\begin{center}
\caption{Properties of the VANDELS-UDS radio selected sample. AGN are classified  according to equation (1). }
\begin{tabular}{llllll}
 \hline
\hline
 & on area& with id & with zspec & with X-ray counterpart & with X-AGN counterpart\\
\hline
All & 114& 96& 16& 27 & 18\\
AGN & 57 & 57& 8 &11& 9\\
\end{tabular}
\end{center}
\end{table*}


\begin{table*}
\begin{center}
\caption{Properties of the VANDELS-UDS X-ray selected sample. Sources are classified as AGN if $L_X\ge10^{42.5}$ erg sec$^{-1}$. }
\begin{tabular}{llllll}
 \hline
\hline
 & on area& with id & with zspec& with radio counterpart & with radio-AGN counterpart\\
\hline
All & 628& 414& 97& 27&11\\
AGN & 345 & 345& 86&18&9\\
\end{tabular}
\end{center}
\end{table*}


\begin{figure}
\includegraphics[scale=0.4]{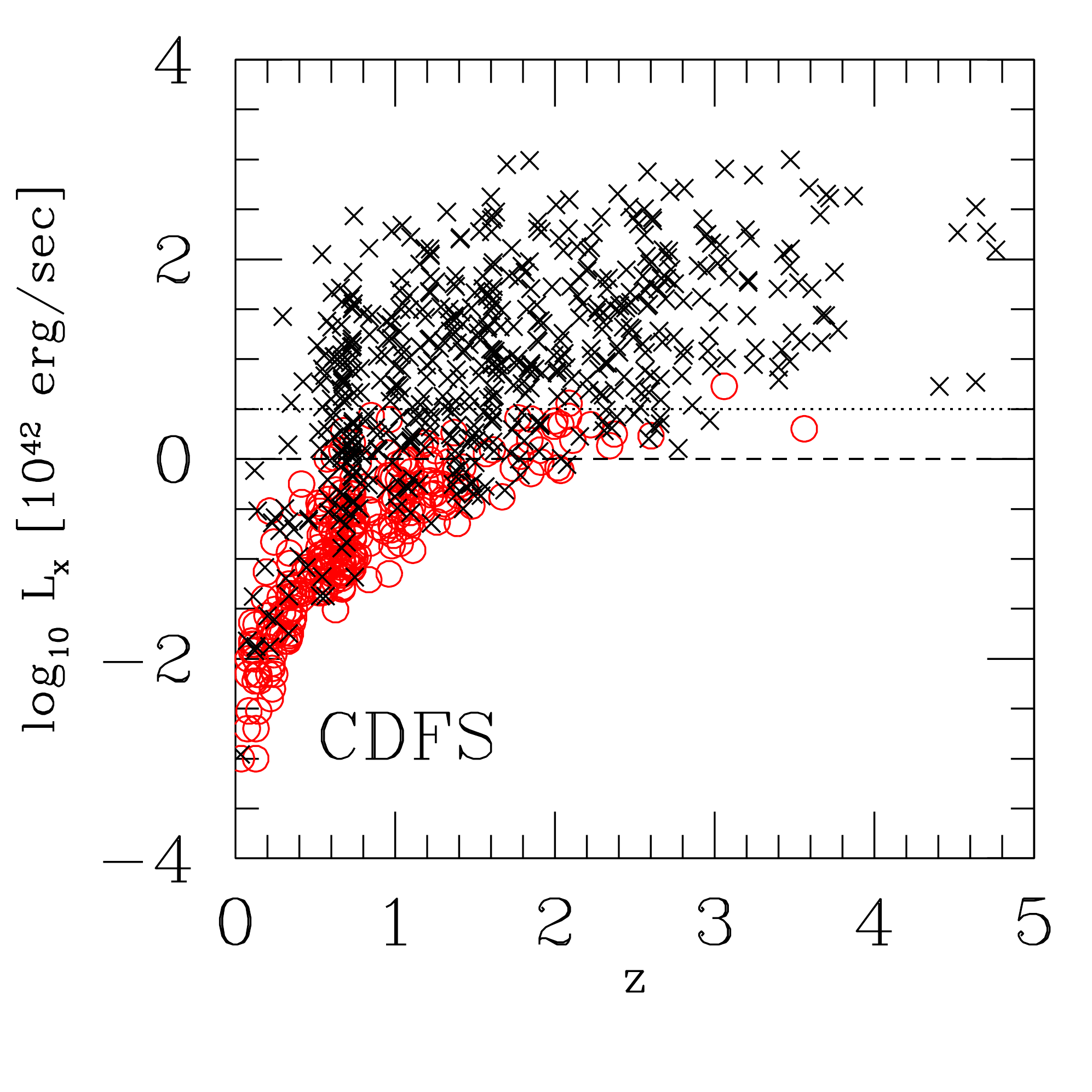}
\caption{Distribution of unobscured X-ray luminosities for those CDFS sources from the Luo et al. (2017) sample with a VANDELS counterpart. AGN are represented by crosses, galaxies by open circles. The dashed line shows the luminosity limit $L_X=10^{42}$ erg sec$^{-1}$ for completeness up to $z\sim 3$, while the dotted line indicates the luminosity threshold  $L_X=10^{42.5}$ erg sec$^{-1}$ above which only AGN are found. AGN and galaxies are classified according to the work of Luo et al. (2017) (see also \S 4.2).}
\end{figure}	

 Deep, 1.4 GHz, radio observations of the whole Extended Chandra Deep Field South (ECDFS) are presented in Miller et al. (2013). This survey covers about a third of a square degree and images 883 radio sources -- out of which 17 are likely multiple-component objects -- at the 5$\sigma$ level down to a peak rms sensitivity of 6 $\mu$Jy.

A smaller area of the ECDFS corresponding to what is called the GOODS Southern Field or the Chandra Deep Field South is covered by VANDELS observations.
We then cross-correlated the  Miller et al. (2013) radio catalogue with the VANDELS-CDFS one in order to provide the 1.4 GHz-selected radio sources with a redshift determination. After correcting for a systematic offset ($\Delta\alpha=0.18^{\prime\prime}, \Delta\delta=0.3^{\prime\prime}$) between radio and VANDELS positions, we chose a maximum tolerance radius of $0.8^{\prime\prime}$, found as the best compromise to maximize the number of true matches while minimizing that of spurious associations (between $\sim 8$\% and $\sim 9$\% in all the cases treated in this work, whereby the percentages of expected mismatches were estimated by vertically offsetting the positions of VANDELS sources by 1$^\prime$). 355 radio objects out of the 883 included in the Miller et al. (2013) work belong to the region covered by the VANDELS MASTER catalogues (cfr \S 2), represented by the shaded cyan area in Figure 1. 
Out of these 355, 277 are found to have a VANDELS counterpart and a photometric redshift estimate, while 186 also possess a spectroscopic redshift determination (cfr Table 1).

Extremely deep X-ray observations of the CDFS are presented in Luo et al. (2017). 
These authors  provide an X-ray catalogue for the 7 Ms exposure of the Chandra Deep Field South which covers a total area of 484.2 arcmin$^2$. The catalogue includes 1055 objects that are detected in up to three X-ray bands: 0.5-7.0 {\it KeV} down to $\sim 1.9 \times 10^{-17}$ erg cm$^{-2}$ sec$^{-1}$, 0.5-2.0 {\it KeV} down to $\sim 6.4 \times 10^{-18}$ erg cm$^{-2}$ sec$^{-1}$ and 2-7 KeV down to $\sim 2.7\times 10^{-17}$ erg cm$^{-2}$ sec$^{-1}$. The projected distribution of these sources onto the VANDELS footprint is represented by the squares in Figure 1. 
Note that, by using ancillary multi-wavelength information, Luo et al. (2017) were able to classify the objects into AGN (711) and star-forming galaxies (344). 
  
Most of the Luo et al. (2017) sources fall in the field covered by VANDELS. In more detail, the number of X-ray objects within the VANDELS area is 996. By proceeding in the same way as for the radio associations,  we found that 
839 of these possess a counterpart and a photometric redshift estimate from the VANDELS-CDFS catalogue, and 650 of them are also endowed with a spectroscopic redshift determination (cfr Table 2). We remind that for the purposes of our work, whenever possible we use spectroscopic redshifts, and only in their absence we rely on photometric estimates. In the relatively few cases when the redshifts provided by the Luo et al. (2017) work did not agree with those from VANDELS, we have also re-calculated the X-ray luminosities of the sources to the values of VANDELS redshifts, starting from the original $L_X$ and assuming an average spectral slope  of 0.8. 

We stress that the lower number of associations found in this work as compared to that of Luo et al. (2017) derives from the fact that 
we only looked for counterparts within VANDELS, without enlarging our search to all catalogues available in the literature. 
This is because we are interested in a homogeneous sample of sources with reliable mass estimates, rather than a collection of redshifts. 
However, note that if we restrict to the region covered by HST observations, we find that the percentages of associations increase to 92\% both in the case of radio-selected and X-ray selected objects. Also, as a sanity check, we have looked at the 709 X-ray sources within the CANDELS region that have a counterpart both in the Luo et al. (2017) and in the VANDELS catalogues. 
In fact, as already stressed at the beginning of this Section, counterparts to sources in our work have been found via a simple matching procedure based on source positioning. More sophisticated methods which use Maximum Likelihood analyses and/or Bayesian techniques (e.g. Salvato et al. 2018) have been recently introduced, and the Luo et al. (2017) counterparts are indeed obtained by following a likelihood-ratio technique presented in Luo et al. (2010).  We find that the percentage of counterparts which coincide in our set and in that of Luo et al. (2017) corresponds to 89\% of the parent catalogue. This figure is very close to the expected fraction of real associations ($\sim 92$\%) previously estimated for our catalogue and provides a further check on the goodness of our matching procedure.


 
 \begin{figure}
\includegraphics[scale=0.4]{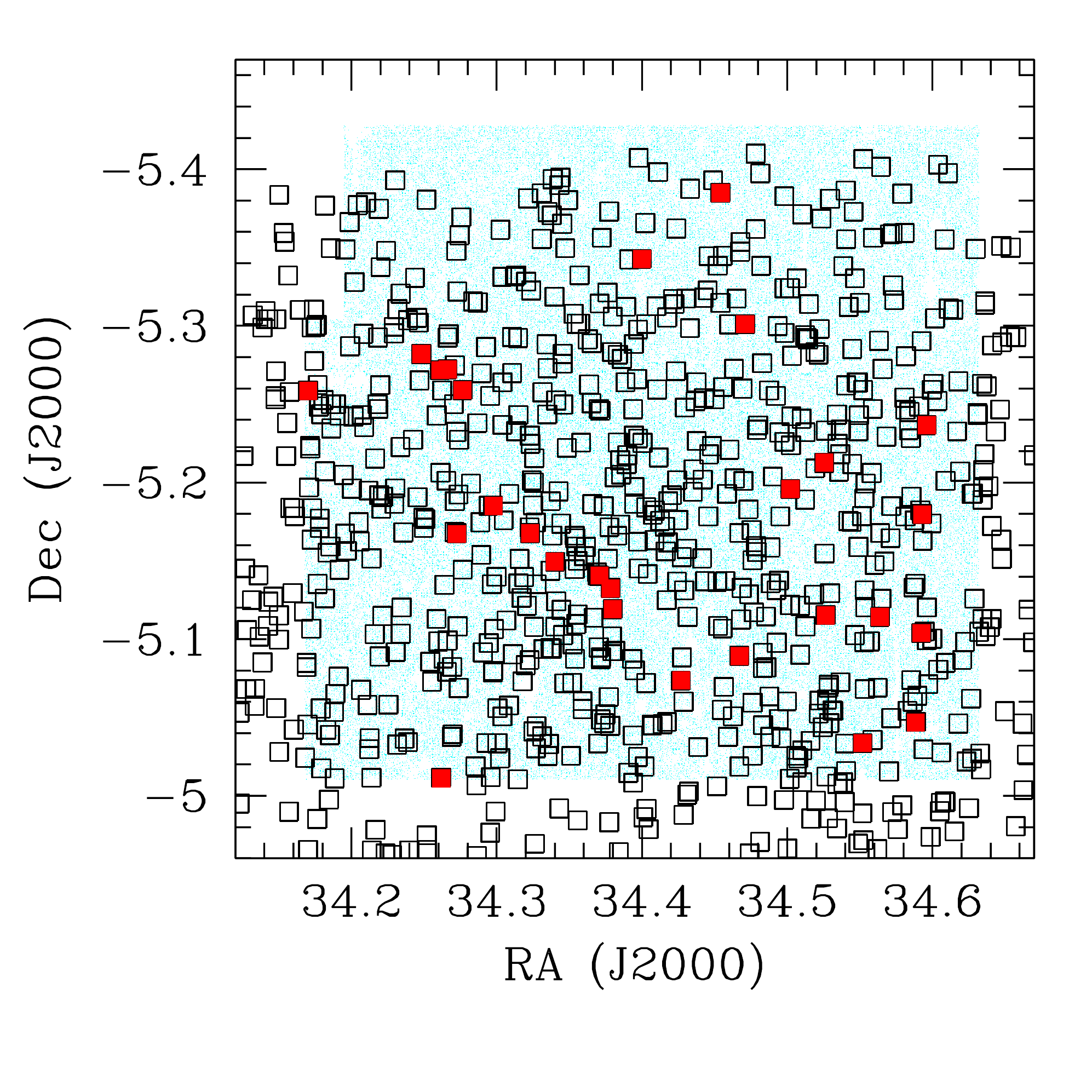}
\caption{Projected distribution onto the sky of X-ray detected sources from the UDS. Filled squares show those objects with a 1.4 GHz counterpart. The shaded cyan  area indicates the photometric coverage of the VANDELS MASTER UDS catalogues (cft text for details).}
\end{figure}	

\begin{figure}
\includegraphics[scale=0.4]{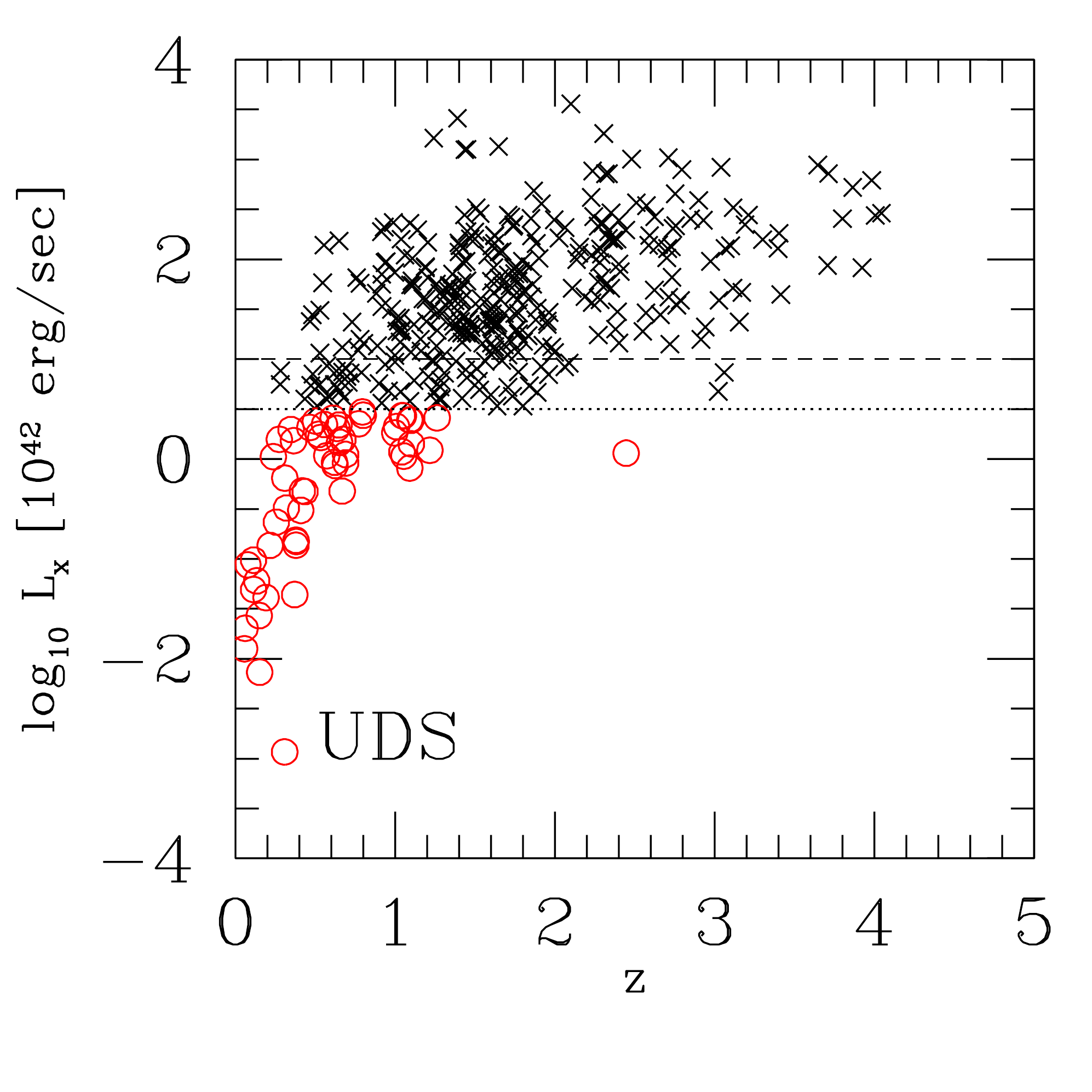}
\caption{Distribution of unobscured X-ray luminosities for those UDS sources from the Kocevski et al. (2018) sample with a VANDELS identification. 
The dashed line shows the luminosity threshold $L_X=10^{43}$ erg sec$^{-1}$ for completeness up to $z\sim 3$, while the dotted line indicates the luminosity limit $L_X=10^{42.5}$ erg sec$^{-1}$ for selection of {\it bona-fide} AGN. Crosses indicate AGN, open circles galaxies selected as specified above and in \S 4.2.}
\end{figure}

Figure 2 illustrates the distribution of unobscured X-ray luminosities for those 7 Ms CDFS sources with a VANDELS identification. The division into classes is based on the work of Luo et al. (2017). AGN are represented by crosses, while galaxies by circles. The plot suggests a survey luminosity limit of the order of $L_X= 10^{42}$ erg sec$^{-1}$ out to $z\sim 3$. However we note that, differently from the radio band,  X-ray bands are sensitive to obscuration in a redshift-dependent way. As a consequence, X-ray selected samples become incomplete at high levels of obscurations in a redshift-dependent manner.\footnote{The incompleteness is expected to decrease with redshift, especially at high X-ray luminosities, while no dependence should arise from host galaxy mass and/or star-formation rate (e.g. Merloni et al. 2014).} As a first order, we can assume that our sample is complete with respect to Compton-thin AGN ($N_H<10^{22}$ cm$^{-2}$) up to $z\sim 3$.
The $L_X= 10^{42}$ erg sec$^{-1}$ luminosity limit is shown by the horizontal dashed line. The horizontal dotted line instead marks the luminosity transition value of $L_X=10^{42.5}$ erg sec$^{-1}$, above which only AGN are found.

In order to investigate the combined properties  of AGN in the radio and X-ray bands, we also looked for X-ray counterparts from the Luo et al. (2017) sample to radio-selected sources. Once again after correcting for a systematic offset between radio and X-ray positions ($\Delta\alpha=0.18^{\prime\prime}, \Delta\delta=0.3^{\prime\prime}$),  we searched for counterparts within a tolerance radius of 2$^{\prime\prime}$. The resulting number of radio sources associated with X-ray emission within the CDFS area covered by VANDELS  is 159 (cfr Tables 1 and 2), which corresponds to $\sim 44$\% of the parent radio population and to $\sim 16$\% of the parent X-ray population. These  are represented by the filled squares in Figure 1.  Chances of spurious associations estimated by vertically shifting the radio positions by 1$^\prime$ are around 1\%. 145 of them also have a counterpart in the VANDELS database.

\subsection{UDS}


\begin{figure*}
\includegraphics[scale=0.4]{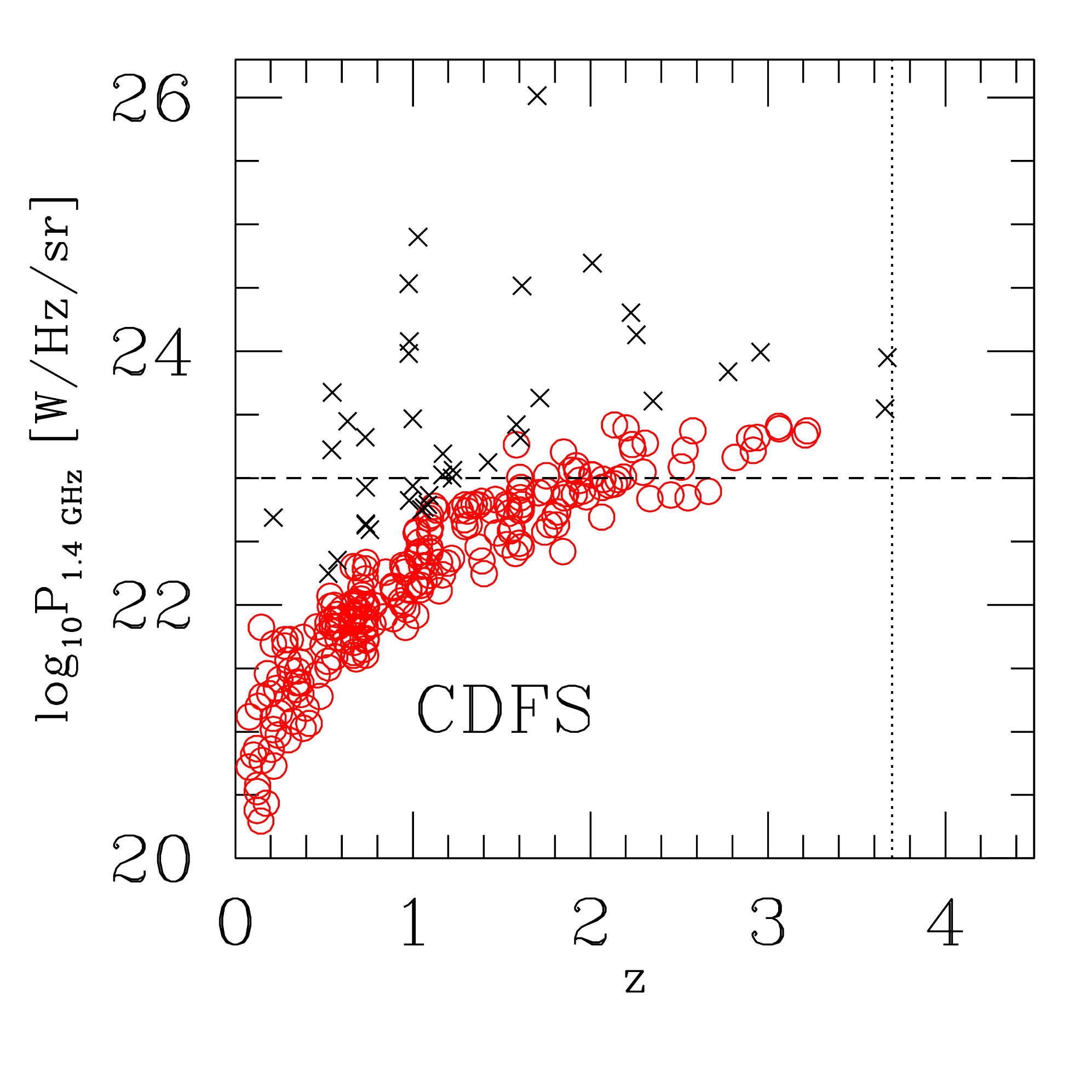}
\includegraphics[scale=0.4]{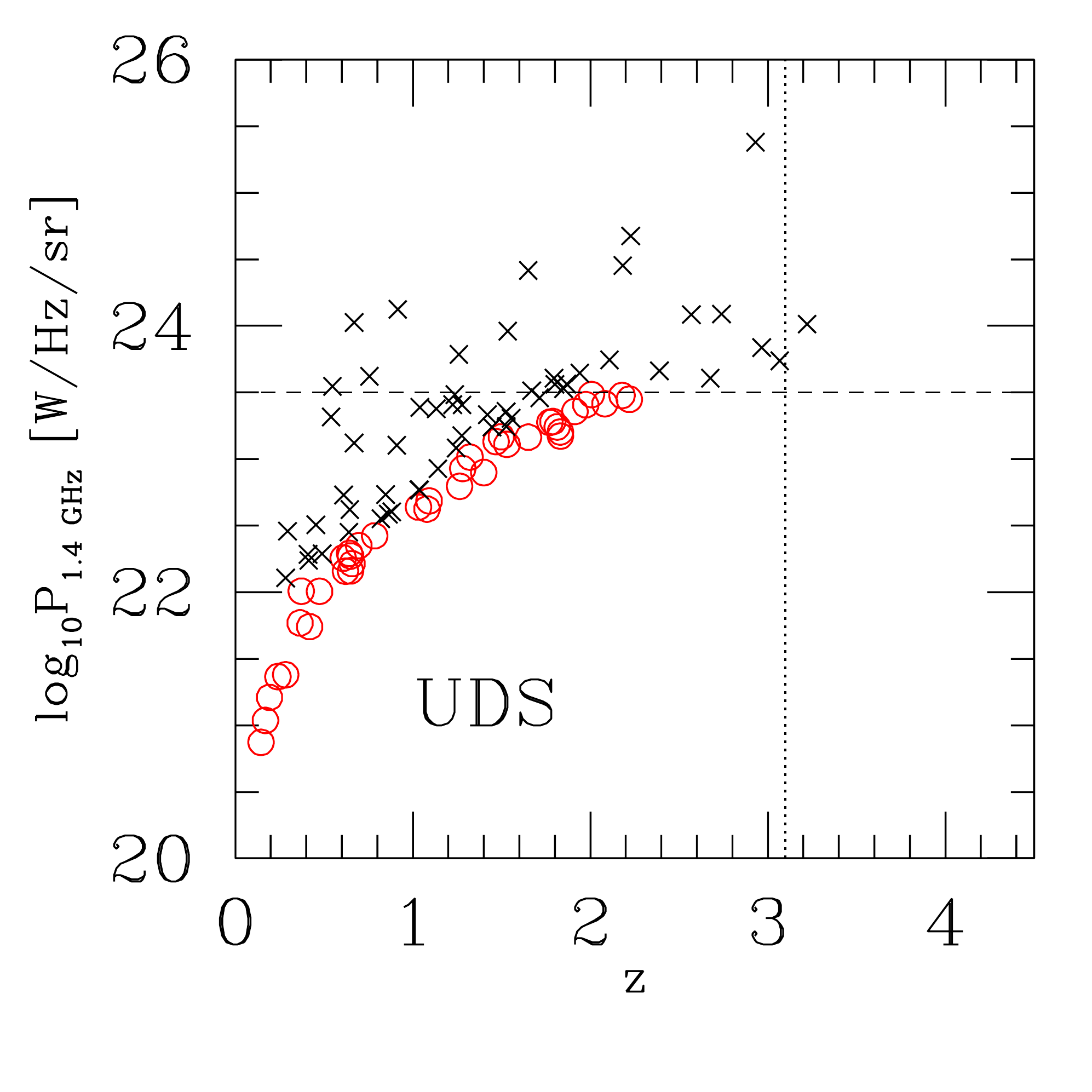}
\caption{Distribution of radio luminosities for  CDFS sources from the Miller et al. (2013) sample (left-hand panel) and UDS sources from the Simpson et al. (2006) sample (right-hand panel) which fall within the VANDELS area. AGN selected with the method presented in \S 4.1 are indicated by crosses, star-forming galaxies by open circles. The dashed horizontal lines illustrate the luminosity thresholds of the two surveys for completeness up to $z\sim 3$, while the dotted vertical ones mark the redshift limits for completeness in the AGN samples (cfr text for details).}
\end{figure*}

Radio observations of the Subaru-XMM Deep/UKIDSS Ultra Deep Survey (hereafter UDS) region are presented by Simpson et al. (2006). These authors provide a catalogue of 505 sources with 1.4 GHz peak radio flux densities greater than 100 $\mu$Jy over an area of 0.81 deg$^{2}$. As Table 3 shows, 114 of such sources are located  within the area covered by the VANDELS  MASTER catalogues and illustrated in Figure 3 by the shaded cyan area. By repeating the same steps as in \S 3.1 and once again correcting for a systematic offset ($\Delta\alpha=0.1^{\prime\prime}, \Delta\delta=0.35^{\prime\prime}$) between radio and VANDELS positions, we find that 96 of them have a counterpart and therefore a photometric redshift estimate from the VANDELS-UDS catalogue, while 16  also possess a spectroscopic redshift determination. 

Relatively deep X-ray observations of the UDS field are presented by the recent work of Kocevski et al. (2018) who provide a catalogue of 864 sources observed down to a depth of $\sim 600$ Ks in the central 100 arcmin$^2$ and to $\sim$ 200 Ks in the remainder of the field. As it is possible to appreciate from Table 4, of these 864 sources 628 fall within the VANDELS area, 414 have a VANDELS counterpart obtained as in \S 3.1, and 97 also possess a spectroscopic redshift determination. The projected distribution of UDS X-ray sources onto the VANDELS footprint is shown by the squares in Figure 3. Also in this case, we stress that we use spectroscopic redshifts whenever possible and photometric redshifts otherwise. As it was for the CDFS, in the  few cases when the redshifts provided by Kocevski et al. (2018)  did not agree with those from VANDELS, we have re-calculated the X-ray luminosities of the sources to the values of VANDELS redshifts, starting from the original $L_X$  and assuming an average spectral slope of 0.8. 
Also, we remind that the lower number of associations found in this work as compared to that of Kocevski et al. (2018) derives from the fact that 
we only looked for counterparts within VANDELS, without enlarging our search to all catalogues available in the literature. At variance with \S3.1, we could not check for how many of the counterparts provided by Kocevski et al. (2018) coincide with those found in the present work, since the Kocevski et al. (2018) catalogues only include redshift information without listing their coordinates. 

Figure 4 illustrates the distribution of unobscured X-ray luminosities for all the Kocevski et al. (2018) sources with a  VANDELS identification. In this case the plot indicates that the survey luminosity limit which ensures completeness up to $z\sim 3$ in the case of Compton-thin ($N_H< 10^{22}$ cm$^{-2}$) sources is  $L_X\sim 10^{43}$ erg sec$^{-1}$. This is shown by the horizontal dashed line. As it was in Figure 2, the horizontal dotted line instead marks the luminosity transition value of $L_X=10^{42.5}$ erg sec$^{-1}$, above which only {\it bona-fide} AGN are selected.
 
As a further step, once again aimed at investigating the combined properties  of AGN in the radio and X-ray bands, after correcting for a systematic offset between X-ray and radio positions ($\Delta\alpha=0.1^{\prime\prime}, \Delta\delta=0.35^{\prime\prime}$), we looked for X-ray counterparts to radio-detected sources within a tolerance radius of 2$^{\prime\prime}$. The resulting number of radio sources within the UDS area covered by VANDELS  which also present X-ray emission from the Kocevski et al. (2018) work is 27 (cfr Tables 3 and 4), out of which 21 have a counterpart in the VANDELS MASTER catalogues. This corresponds to $\sim 4$\% of the parent X-ray population and to $\sim 24$\% of the parent radio population. As it will be discussed at greater length in the following Sections, we stress that the discrepancy between the number of radio counterparts to X-ray sources as obtained in the CDFS and in the UDS fields is mainly to be attributed to the limiting fluxes of the radio and X-ray surveys which in both cases are much brighter in the UDS with respect to the CDFS.
 
  
 
 \section{AGN selection}
 \subsection{Radio}
 A tricky issue one usually encounters when dealing with monochromatic radio surveys is the assessment of whether the radio signal produced by a generic source stems from the activity of a central AGN or rather from star-formation processes ongoing within the host galaxy. One possible way to distinguish between these two phenomena is via the excess of radio emission with respect to that expected from pure star-forming processes (e.g. Padovani et al. 2015; Bonzini et al. 2015; Smolcic et al. 2017; Ceraj et al. 2018). This method is proven to be quite efficient. However since it requires deep multi-wavelength information, its application is necessarily restricted to only well studied deep fields such as COSMOS or CDFS.
 
In order to overcome this issue, Magliocchetti et al. (2014) then introduced a different approach  which only requires knowledge of the radio luminosity of the sources under exam. Its successful applications to a number of cases were presented in Magliocchetti et al. (2016, 2017, 2018 and 2018b).

 Briefly, it  is based on the  results of McAlpine, Jarvis \& Bonfield (2013) who used the optical and near infrared Spectral Energy Distributions of a sample of 942 radio sources  (out of 1054 objects selected at 1.4 GHz, with a completeness level of 91\%) from the VIDEO-XMM3 field to distinguish between star forming and AGN-powered galaxies and derive their redshifts. 
These authors provide luminosity functions for the two classes of sources up to redshifts $\sim 2.5$ and find different evolutionary behaviours, with star-forming objects evolving in luminosity in a much stronger way ($\propto (1+z)^{2.5})$ than radio-selected AGN ($\propto (1+z)^{1.2}$).

Investigations of their results show that the radio luminosity P$_{\rm cross}$ beyond which  AGN-powered galaxies become the dominant radio population scales with redshift roughly as
\begin{eqnarray}
\rm Log_{10}P_{\rm cross}(z)=\rm Log_{10}P_{0,\rm cross}+z,
\label{eq:P}
\end{eqnarray}
at least up to $z\sim 1.8$. $P_{0,\rm cross}=10^{21.7}$[W Hz$^{-1}$ sr$^{-1}$] is the value which is found in the local universe and which roughly coincides with the break in the radio luminosity function of star-forming galaxies (cfr Magliocchetti et al. 2002; Mauch \& Sadler 2007). Beyond this value,  their  luminosity function steeply declines, and the contribution of star-forming galaxies to the total radio population is drastically reduced to a negligible percentage. The same trend is true at higher redshifts, and since the radio luminosity function of star-forming galaxies drops off in a much steeper way than that of AGN at all $z$, we expect the chances of contamination between these two populations for $P>P_{\rm cross}$ to be rather low (around $\sim$ 15\% at $P_{\rm cross}$, rapidly dropping to $\sim$ 5\% already at $2\cdot P_{\rm cross}$ -- see the Appendix of Magliocchetti et al. 2018 for a thoroughful investigation of possible contamination issues).

\begin{figure*}
\includegraphics[scale=0.4]{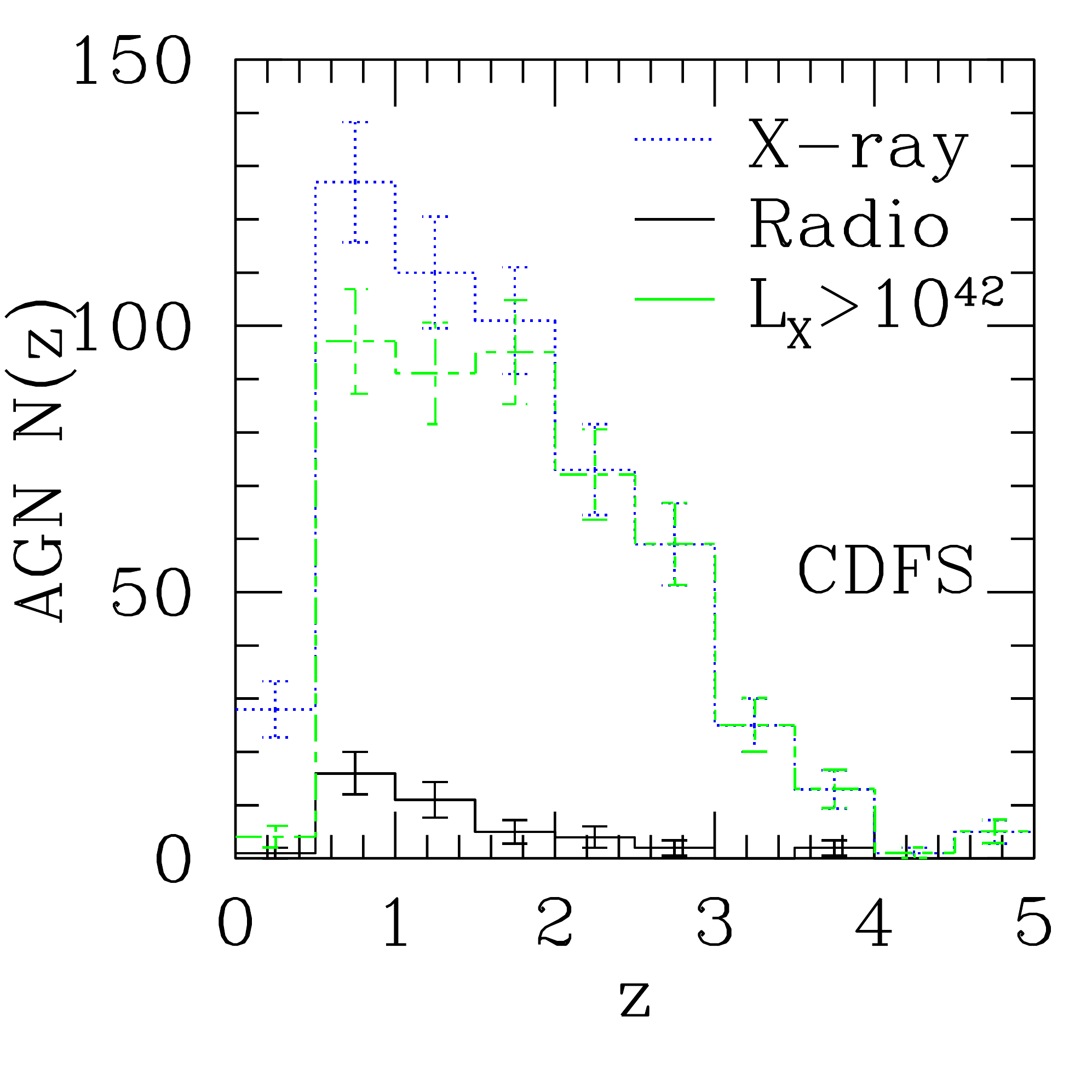}
\includegraphics[scale=0.4]{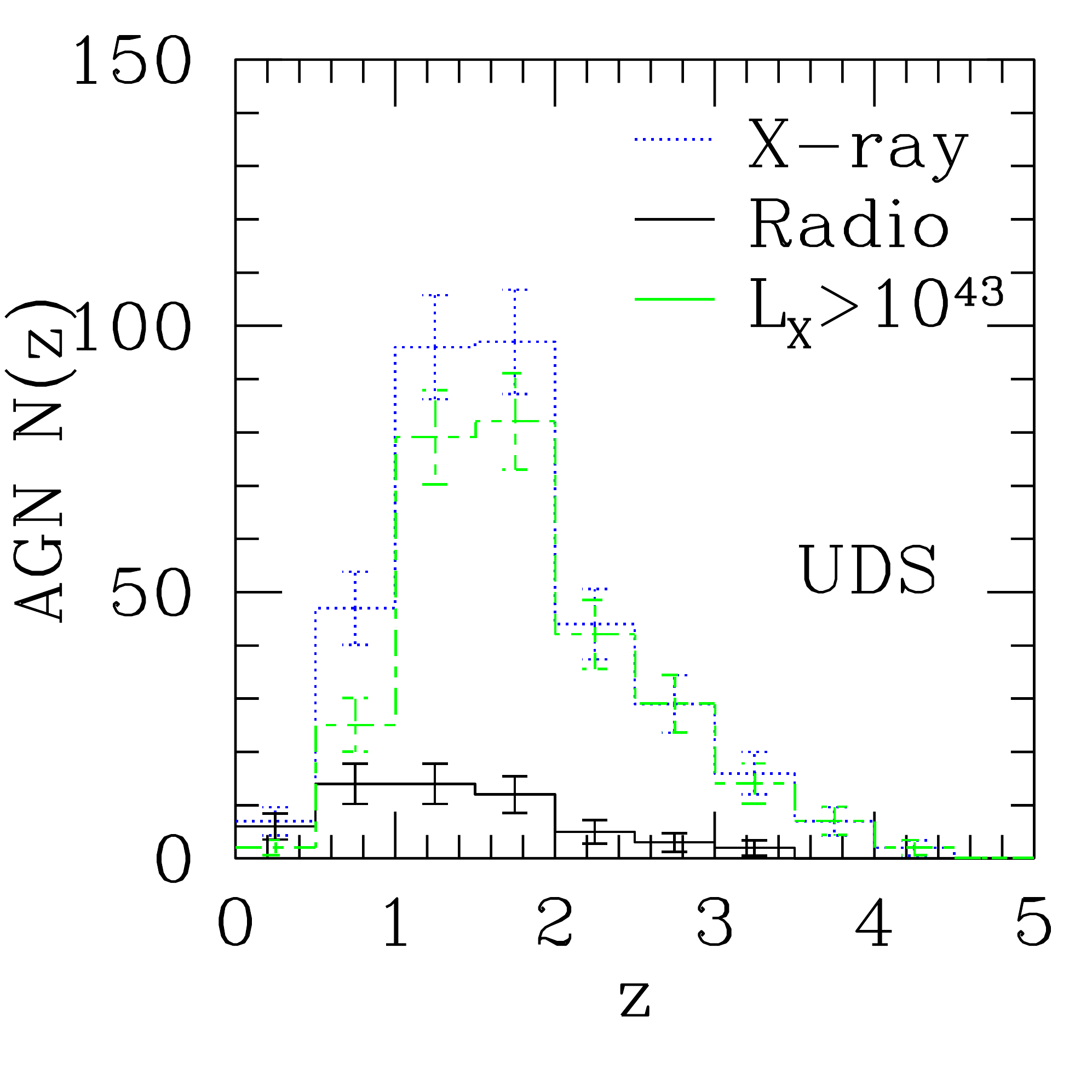}
\caption{Redshift distributions for AGN in the CDFS (left-hand panel) and UDS (right-hand panel). The solid histograms show those obtained for radio-selected AGN,  the dotted histograms those for the two X-ray surveys, while the long-short dashed histograms indicate the redshift distributions derived in the case of X-ray samples which are complete up to $z\sim 3$. These correspond to  luminosities $L_X\ge 10^{42}$ erg sec$^{-1}$ for CDFS and $L_X\ge 10^{43}$ erg sec$^{-1}$ for UDS (cfr \S 3 and Figures 2 and 4). Error-bars represent 1$\sigma$ Poisson uncertainties. }
\end{figure*}

 \begin{figure}
\includegraphics[scale=0.45]{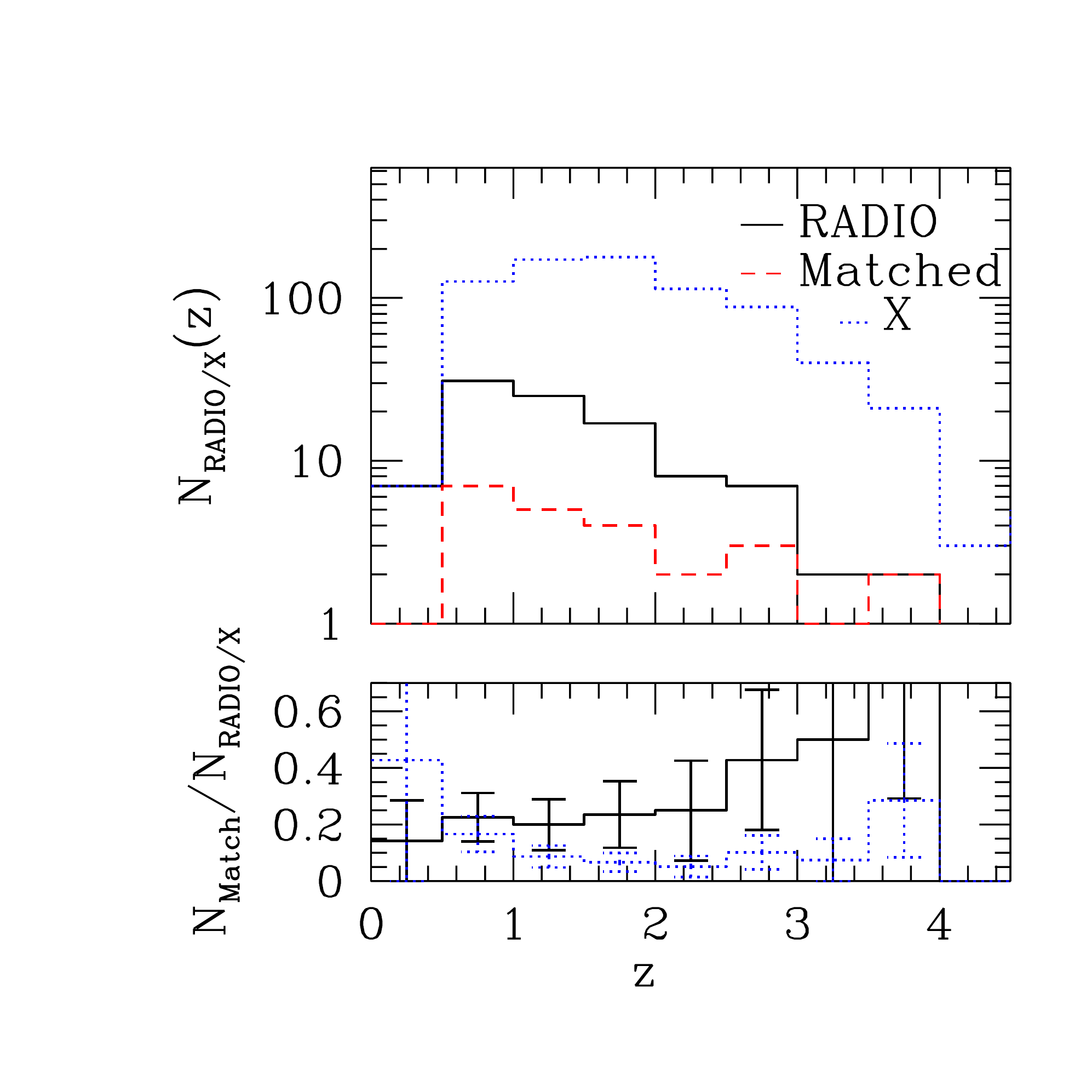}
\caption{Top panel: redshift distribution for the combined sample of CDFS+UDS AGN selected in the X-ray above the completeness levels discussed in \S 3 and \S 4 (dotted line), for all CDFS+UDS AGN selected in the radio band (solid line) and for those sources which are identified as AGN both at radio and X-ray wavelengths (dashed line). Bottom panel: ratio between number of AGN detected at both radio and X-ray wavelengths and number of X-ray selected AGN (dotted line, multiplied by a factor 3 for the sake of clarity) and radio-selected AGN (solid line) as a function of $z$. Error-bars represent 1$\sigma$ Poisson uncertainties. }
\end{figure}

Radio luminosities to be plugged in equation (1) have been obtained according to the relation:
\begin{eqnarray}
\rm P_{1.4 \rm GHz}=\rm F_{1.4 \rm GHz} D^2 (1+z)^{3+\alpha},
\end{eqnarray}
where the result is in [W Hz$^{-1}$ sr$^{-1}$] units, $F_{\rm 1.4 GHz}$ is the radio flux at 1.4 GHz, D is the angular diameter distance and $\alpha$ is the spectral index of the radio emission ($\rm F(\nu)\propto \nu^{-\alpha}$). \\
As we do not have estimates of the quantity $\alpha$ for each source, we adopted the average value $\alpha=0.7$ found for similar surveys (e.g. Randall et al. 2012 and references therein) both for star-forming galaxies and for AGN emission, in agreement with results from  Magliocchetti et al. (2016) who find $<\alpha>\simeq 0.685$ -- independent of flux and redshift -- for radio sources observed in the Lockman Hole.

We then distinguished between AGN-powered galaxies and star-forming galaxies by means of equation (\ref{eq:P}) for $z\le 1.8$ and by fixing $\rm Log_{10}P_{\rm cross}(z)=23.5$ [W Hz$^{-1}$ sr$^{-1}$ ] at higher redshifts (cfr McAlpine, Jarvis \& Bonfield 2013). 
By doing this, we end up with 43 radio-emitting AGN (out of which 21 have a spectroscopic redshift determination, cfr Table 1) in the CDFS and 57 radio-emitting AGN (out of which 8 have a spectroscopic redshift determination, cfr Table 3) in the UDS. 

As shown by the vertical dotted lines in Figure 5 which presents the 1.4 GHz luminosities of radio-detected sources in the CDFS (left-hand panel) and UDS (right-hand panel), whereby AGN are indicated by crosses and star-forming galaxies by open circles, the relative depths of the CDFS and UDS radio surveys ensure that the samples of AGN selected by following the method previously highlighted  are complete in the former case up to $z\sim 3.7$, while in the case of UDS  up to $z\sim 3.1$. The horizontal dashed lines in each panel of Figure 5 instead represent the completeness levels of the full (AGN+star-forming galaxy) radio surveys up to $z\sim 3$. Since the CDFS radio survey is much deeper than the one carried on the UDS, it is complete down to fainter luminosities. However, despite of this, since in the present work radio-active AGN are purely selected on the basis of radio luminosity, it turns out that for this latter population the two radio surveys have comparable depths and return AGN samples of comparable size and properties.

\subsection{X-ray}
As already specified in \S 3.1, a classification into AGN or galaxies for X-ray sources observed in the CDFS is already provided in the work of Luo et al. (2017). 
As summarized in the second line of Table 2, there are 680 X-ray selected AGN from the Luo et al. (2017) work which fall within the area covered by VANDELS. Out of these, 562 have a counterpart  from the VANDELS database and therefore a photometric redshift estimate, while 389 of them also possess a spectroscopic redshift determination. The 7 Ms survey is complete down to luminosities $L_X=10^{42}$ erg sec$^{-1}$ (cfr Figure 2), and therefore includes all X-ray detected AGN up to redshifts $z\sim 3$.
 
For what concerns X-ray sources observed in the UDS, the work by Kocevski et al. (2018) does not provide any classification. 
Therefore we chose to consider as AGN all those objects with an X-ray luminosity greater than 10$^{42.5}$ erg sec$^{-1}$ (cfr Luo et al. 2017). The chosen threshold ensures that only {\it bona-fide} AGN will be considered in the following analysis and that our UDS AGN sample is unaffected by X-ray emitting star-forming interlopers. However, we should bear in mind that we might miss real AGN with X-ray luminosities fainter than the chosen 10$^{42.5}$ erg sec$^{-1}$ threshold which are instead included in the CDFS by the classification  of Luo et al. (2017). 
Furthermore, while completeness up to $z\sim 3$ in the CDFS is reached for $L_X\simgt 10^{42}$ erg sec$^{-1}$, in the UDS it is only achieved at a luminosity level  which is ten times higher, $L_X\simeq 10^{43}$ erg sec$^{-1}$ (cfr Figure 4). It follows that the AGN sample originating from the Kocevski et al. (2018) survey is pure, but only complete for luminosities above this value, as in the range $10^{42.5}$ erg sec$^{-1}$ $\simlt L_X \simlt$ $10^{43}$ erg sec$^{-1}$ we are missing sources with $z\simgt 2$.

Bearing these caveats in mind, we can nevertheless proceed with the chosen $L_X=10^{42.5}$ erg sec$^{-1}$  threshold to select X-ray emitting AGN in the UDS field. As specified in Table 4, by doing so we end up with 345 AGN with a VANDELS counterpart. 86 of them are also provided with a spectroscopic redshift determination. 


\begin{figure*}
\includegraphics[scale=0.4]{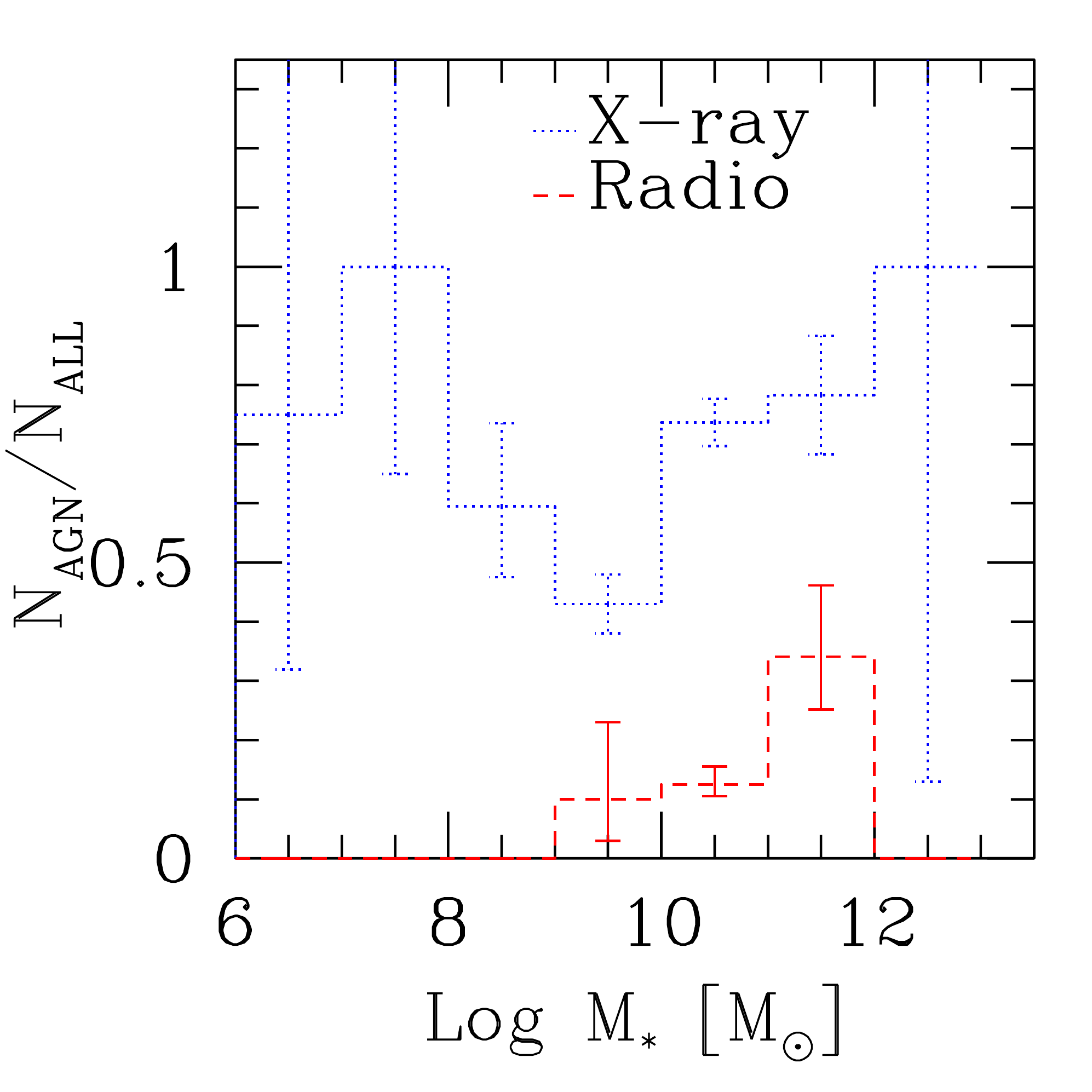}
\includegraphics[scale=0.4]{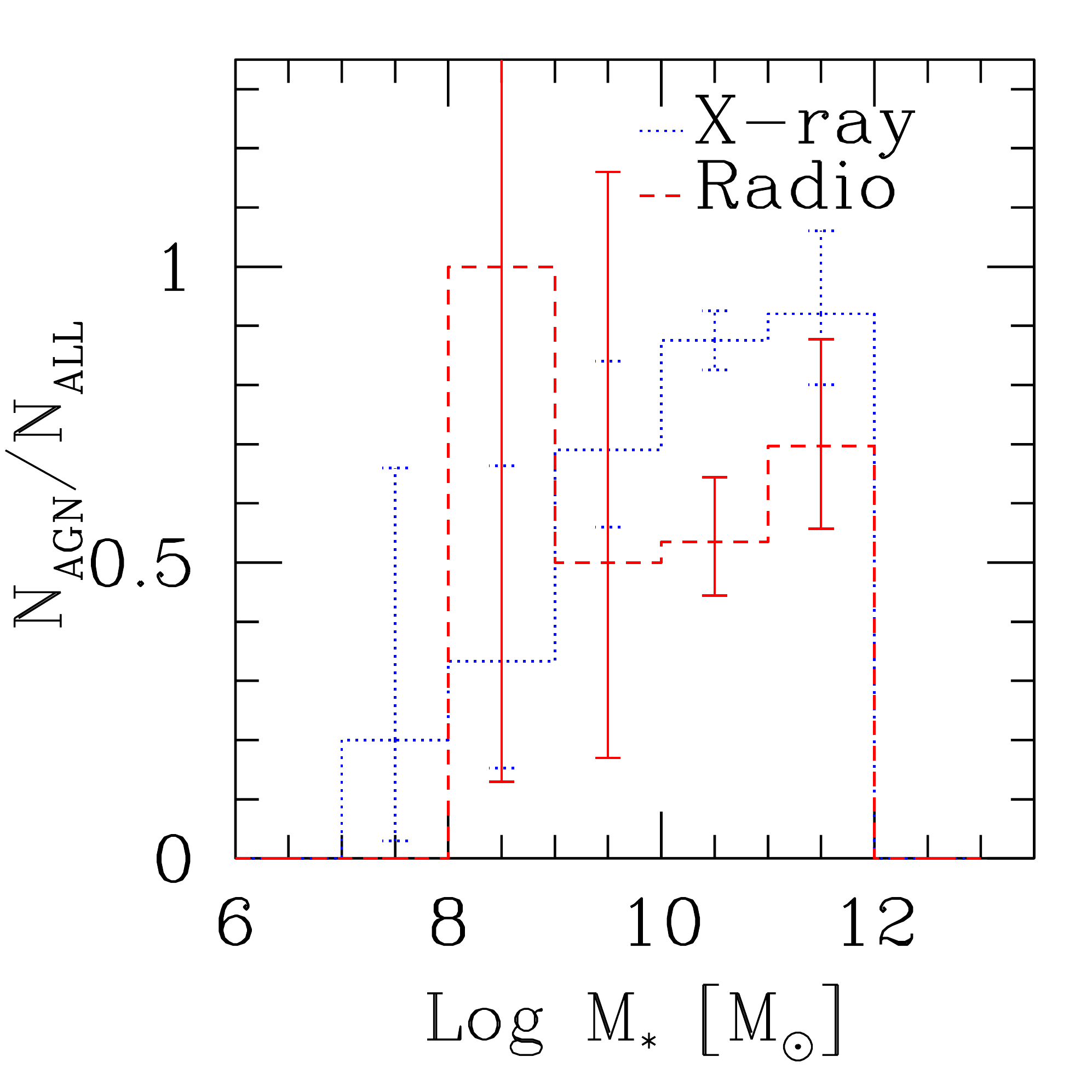}
\caption{Fraction of sources detected either in the X-ray (dotted lines) or radio (dashed lines) wavebands which host an AGN at their center as a function of mass. Error-bars indicate 1$\sigma$ Poisson uncertainties estimated following Gehrels (1986). The left-hand panel refers to the CDFS, while the right-hand panel to UDS. More information is provided in Tables 5 and 6.}
\end{figure*}

\section{Results}

\begin{figure*}
\includegraphics[scale=0.43]{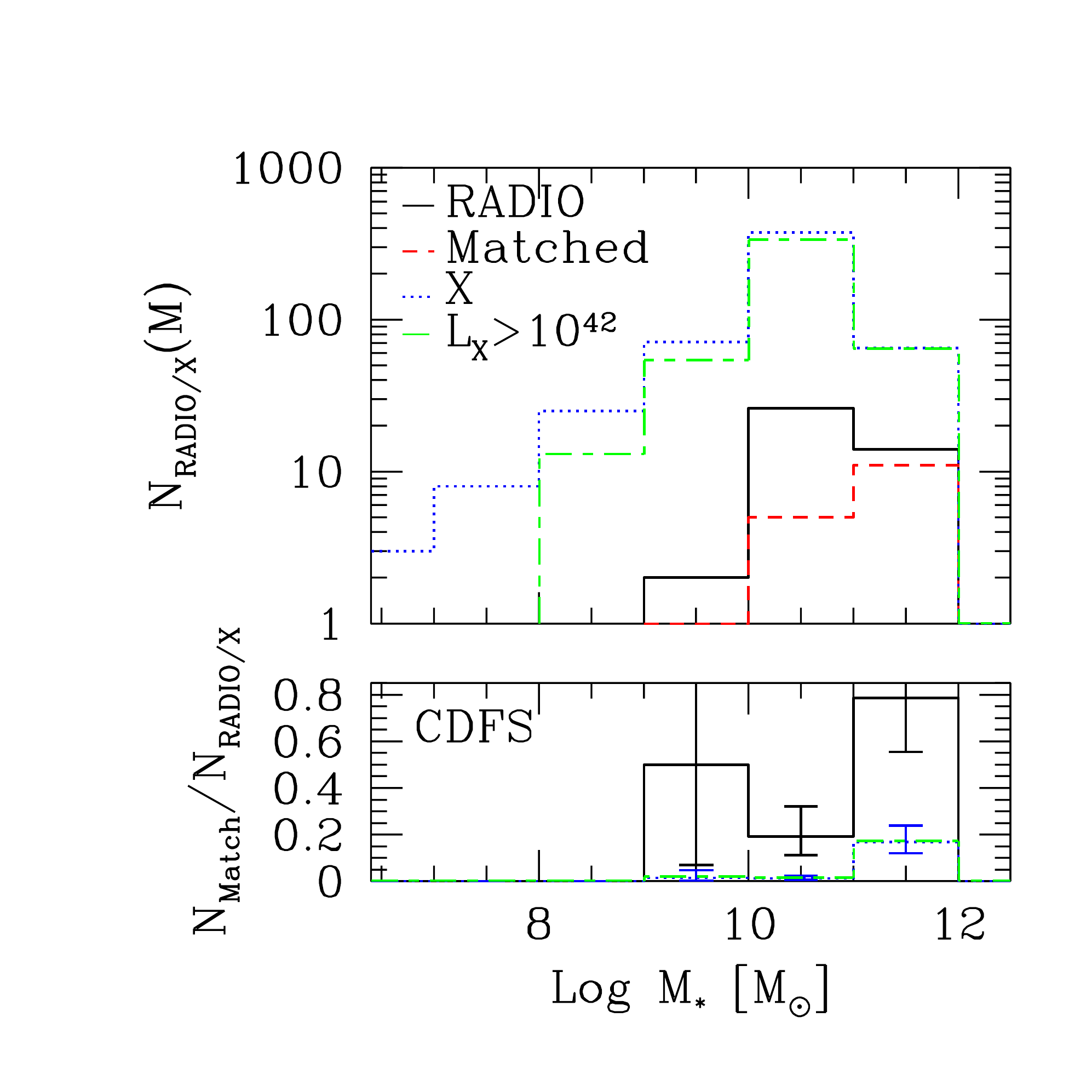}
\includegraphics[scale=0.43]{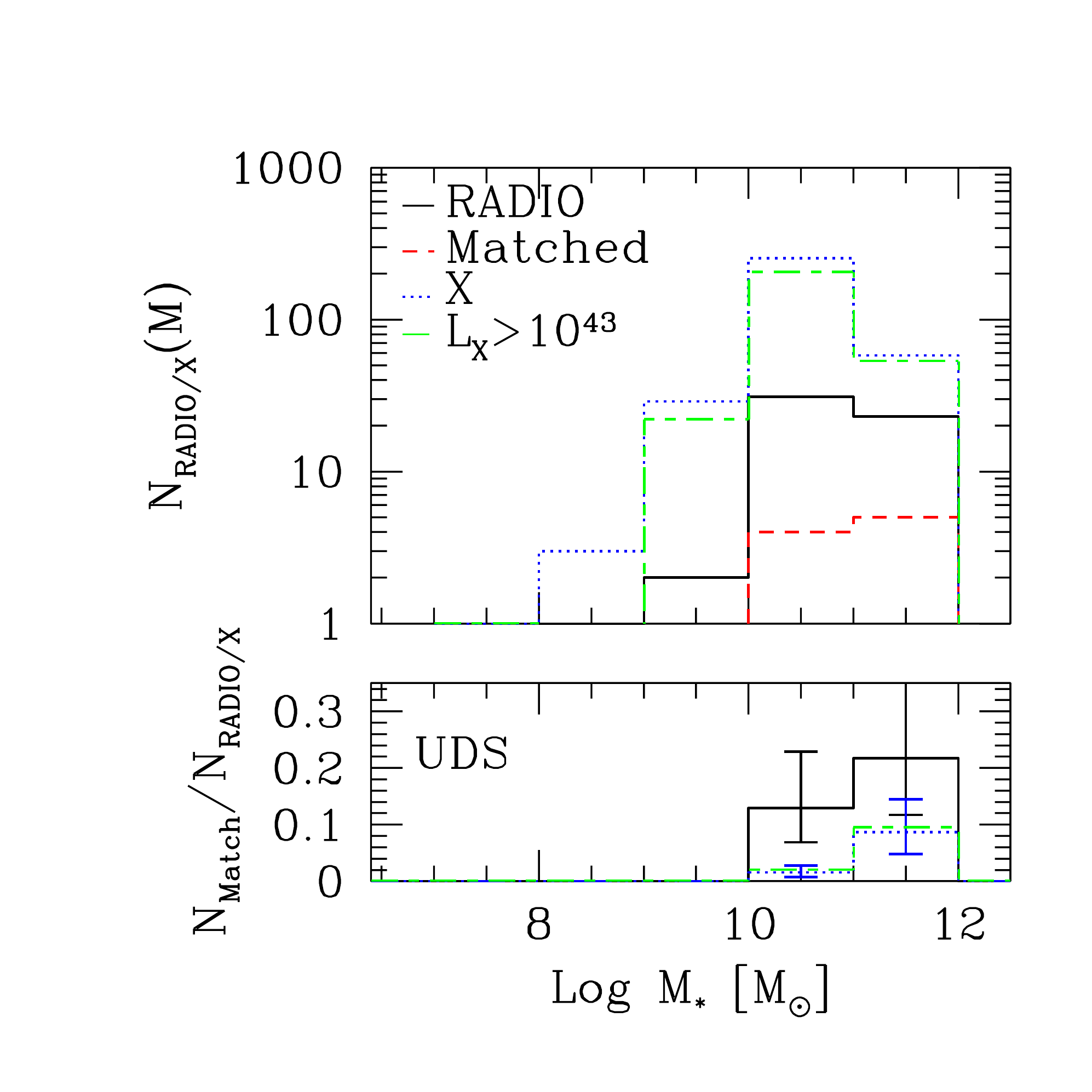}
\caption{Top panels: stellar mass distribution for the hosts of all AGN selected in the X-ray (dotted lines),  all AGN selected in the X-ray above the completeness limits of the surveys in the two fields (long-short dashed lines), all AGN selected in the radio (solid lines) and of those sources which are identified as AGN at both radio and X-ray wavelengths (dashed lines). Bottom panels: ratio between number of AGN simultaneously emitting at radio and X-ray wavelengths and total number of X-ray selected AGN (dotted lines) and radio-selected AGN (solid lines) as a function of stellar mass. Error-bars represent 1$\sigma$ Poisson uncertainties estimated following Gehrels (1986). More information is provided in Tables 5 and 6. The left-hand panel refers to the CDFS, the right-hand one to  UDS.}
\end{figure*}

The final samples which will be used in the following of our work are then two-folded. 
On the one hand we will consider X-ray-selected AGN with available ancillary information in the VANDELS database. These are 562 in the CDFS and 345 in the UDS.  On the other hand, we will also consider radio-selected AGN, again with available ancillary info in the VANDELS database. There are 43 of such sources in the CDFS and 57 in the UDS. The number of objects which are identified as AGN both at X-ray and radio wavelengths is 17 in the CDFS and 9 in the UDS. These all have a VANDELS counterpart. 

On the basis of what discussed in \S 4.1 and \S 4.2, it is important to stress that, in spite of  the different depths, given our adopted definition for radio AGN, the radio surveys performed by Miller et al. (2013) on the CDFS and by Simpson et al. (2006) on the UDS return AGN samples of comparable completeness levels, properties and sizes, at least up to $z\sim 3$. However, this is not true for what concerns the X-ray surveys on the same fields. Indeed, the observations presented by Luo et al. (2017) for the CDFS are much deeper than those presented by Kocevski et al. (2018) for the UDS, and consequently include about a factor 2 more (if one compensates for the larger area covered by VANDELS  on the UDS, cfr Figures 1 and 3) X-ray-selected AGN.  These different completeness levels need to be taken into account in the following analysis as for instance they explain why only 9 radio-selected AGN belonging to the UDS are found to have an X-ray AGN counterpart, while the number is doubled in the CDFS.

\begin{table*}
\begin{center}
\caption{Properties of the CDFS samples. The first column reports the interval for the stellar mass of the surveyed objects, the second column the number of X-ray sources, the third column the number of X-ray AGN, while the fourth column provides the ratio between number of X-ray-emitting AGN and total number of X-ray-selected sources.  The fifth, sixth and seventh columns report the same quantities as columns 2 to 4 but for radio-selected objects, while columns 8 to 10 refer to the number of AGN which are active at both radio and X-ray wavelengths, and to the fractions of such AGN with respect to the total number of X-ray emitting AGN and radio-emitting AGN. The quoted errors correspond to 1$\sigma$ Poisson uncertainties estimated following Gehrels (1986).}
\begin{tabular}{lllllllllll}
 \hline
\hline
Mass Interval [$M_\odot$]&X-ray& AGN$_{\rm X}$& F$_{\rm AGN_X}$ & Radio & AGN$_{\rm R}$ & F$_{\rm AGN_R}$&AGN$_{\rm R+X}$& AGN$_{\rm R+X}$/AGN$_{\rm X}$& AGN$_{\rm R+X}$/AGN$_{\rm R}$\\
\hline
6-7& 4& 3& $0.75^{+0.73}_{-0.43}$& 0&-&-&-&-&-&\\
7-8& 8& 8& $1.0^{+0.49}_{-0.35}$& 1&-&-&-&-&-&\\
8-9&42&25&$0.60^{+0.14}_{-0.12}$&2&-&-&-&-&-&\\
9-10&165& 71&$0.43^{+0.05}_{-0.05}$ &20&$ 2$&$0.10^{+0.13}_{-0.07}$&1&$0.014^{+0.033}_{-0.012}$&$0.5^{+1.16}_{-0.43}$&\\
10-11&506&373&$0.74^{+0.04}_{-0.04}$&207&$26$&$0.13^{+0.03}_{- 0.02}$&5&$0.013^{+0.009}_{-0.006}$&$0.19^{+0.13}_{- 0.08}$&\\
11-12&83& 65&$0.8^{+0.1}_{-0.1}$&41&14&$0.34^{+0.12}_{-0.09}$&11&$0.17^{+0.07}_ {-0.05}$&$0.78^{+0.32}_{-0.23}$&\\
12-13&1& 1&$1.0^{+1.32}_{-0.87}$&-&-&-&-&-&-&\\
\end{tabular}
\end{center}
\end{table*}

\begin{table*}
\begin{center}
\caption{Properties of the UDS samples. Column description as in Table 5.}
\begin{tabular}{lllllllllll}
 \hline
 Mass Interval [$M_\odot$]&X-ray& AGN$_{\rm X}$& F$_{\rm AGN_X}$ & Radio & AGN$_{\rm R}$ & F$_{\rm AGN_R}$&AGN$_{\rm R+X}$& AGN$_{\rm R+X}$/AGN$_{\rm X}$& AGN$_{\rm R+X}$/AGN$_{\rm R}$\\
\hline
\hline
6-7& 1& -&-& -&-&-&-&-&-&\\
7-8& 5& 1& $0.2^{+0.46}_{-0.17}$& -&-&-&-&-&-&\\
8-9&9&3&$0.33^{+0.33}_{-0.18}$&1&1&$1.0^{+2.32}_{-0.87}$&-&-&-&\\
9-10&42& 29&$0.69^{+0.15}_{-0.13}$ &4&$ 2$&$0.50^{+0.66}_{-0.33}$&-&-&-&\\
10-11&289&253&$0.87^{+0.05}_{-0.05}$&53&$31$&$0.58^{+0.11}_{-0.09}$&4&$0.016^{+0.012}_{-0.008}$&$0.13^{+0.10}_{-0.06}$&\\
11-12&63& 58&$0.92^{+0.14}_{- 0.12}$&33&22&$0.67^{+0.18}_{-0.14}$&5&$0.086^{+0.058}_{-0.038}$&$0.23^{+0.15}_{-0.10}$&\\
12-13&-&- &-&-&-&-&-&-&-&\\
\end{tabular}
\end{center}
\end{table*}

The situation can be better appreciated in Figure 6, which shows for the CDFS (left-hand panel) and UDS (right-hand panel) the redshift distributions of the various AGN samples. The solid histograms refer to radio-selected AGN, the dotted histograms to X-ray selected AGN, while the long-short dashed histograms show the redshift distributions of the complete samples of X-ray AGN, corresponding to luminosities $L_X\ge 10^{42}$ erg sec$^{-1}$ in the case of CDFS and $L_X\ge10^{43}$ erg sec$^{-1}$ for UDS. As already anticipated, the plots indicate that, while the radio samples have comparable properties both in terms of size and distributions, the two X-ray AGN datasets coming from CDFS and UDS are rather different,  as the CDFS includes many more low redshift  ($0.5\simlt z\simlt 1$) and high-redshift ($z\simgt 2$) sources than the UDS.
 
In the top panel of Figure 7, we then report the total redshift distributions obtained from the combination of the two fields. Again, the solid histogram represents radio-selected AGN, the dotted histogram X-ray selected AGN coming from the complete samples shown by the short-long dashed lines in Figure 6 and discussed in \S 4.2, and the dashed histogram presents the redshift distribution of those AGN which are detected in the radio and X-ray. Note that since these latter sources all have X-ray luminosities  above the completeness limits of the CDFS and UDS surveys, their numbers are not affected by completeness issues.
The bottom panel of Figure 7 instead presents the ratio between the number of AGN detected at both wavelengths and those detected in the radio (solid histogram) and X-ray (dotted histogram, multiplied by a factor 3 for the sake of clarity) as a function of redshift.

The trends presented in the top panel of Figure 7 show that the distribution of radio-selected AGN starts declining beyond $z\simeq 1$, while that of X-ray emitting AGN has a peak  in the range $1 \simlt z \simlt 2$ and only starts decreasing in the higher redshift universe. Therefore, it appears that the two classes of AGN have somehow different evolutionary behaviours. The distribution of AGN simultaneously emitting at both frequencies instead presents a trend which is intermediate between those of the two parent populations of X-ray and radio AGN emitters, with a peak at $0.5\simlt z\simlt1$ then followed by more shallow decrement in the number of sources than that observed in radio-selected AGN. Within the uncertainties associated to our analysis, and as illustrated in the bottom panel of Figure 7, the above distributions then indicate that the fraction of AGN which emit at both radio and X-ray wavelengths is independent of redshift. In other words, {\it given two underlying distributions of AGN active in the radio or in the X-ray bands, the probability that an AGN will be simultaneously active at both wavelengths is the same at all times}.

\subsection{Mass Distributions}
An exercise similar to that just carried out on the redshift distributions of AGN emitting at X-ray and radio wavelengths can be performed in the case of the mass distribution of the hosts of such sources\footnote{Note that the masses computed in VANDELS  do not include an AGN component in the SED-fitting procedure. This could in principle affect their estimates, even though we do not expect large variations at least in all those cases where the AGN contribution does not dominate over that of the host galaxy in the optical and NIR bands, i.e.  for most radio-active AGN and most X-ray AGN which do not present broad-line components in their spectra. Based on the results of  Yang et al.  (2017) who only found 19 broad-line AGN within their sample of 395 X-ray sources from the Luo et al. (2017) catalogue with counterparts in CANDELS, we expect the contamination level due to such sources to be of the order of 5\%.} The results are presented in Tables 5 and 6 and Figures 8 and 9.

\begin{figure*}
\includegraphics[scale=0.4]{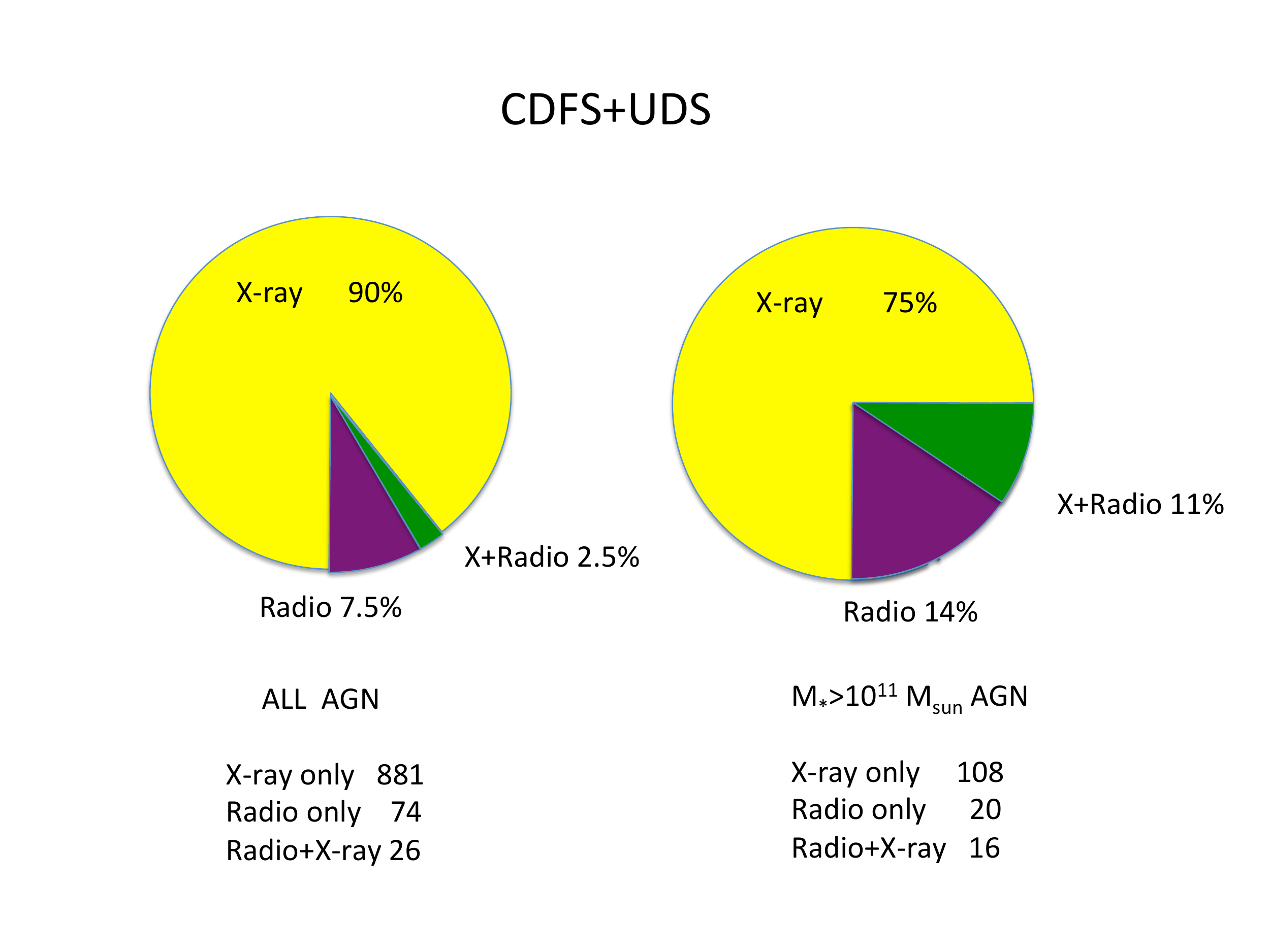}
\caption{Pie chart illustrating the contribution of the different emitters to the total (radio+X-ray) AGN population obtained by combining together the UDS and CDFS fields. The areas shaded in yellow correspond to X-ray emitters, those shaded in purple to radio-emitters while the green regions indicate the contribution from those AGN which simultaneously emit in the radio and X-ray bands. The left-hand panel shows the case for AGN associated with galaxies of all stellar masses, while that on the right-hand side corresponds to AGN residing within hosts with $M_* \ge 10^{11}M_\odot$.}
\end{figure*}	

Figure 8 shows the fraction of X-ray (dotted histograms) and radio-selected (dashed histograms) sources which host an AGN at their center as a function of stellar mass of their hosts. The left-hand plot illustrates the case for CDFS, while the right-hand one is for UDS.  As both panels clearly show, the behavior of the distributions obtained for objects selected at different wavelengths are quite different. Indeed, for radio-emitting sources our data mirrors the well-known result for radio-active AGN to preferentially reside in galaxies of large, $M_*\simgt 10^{10} M_\odot$, stellar content (e.g. Best et al. 2005; Smolcic et al. 2009; Magliocchetti et al. 2014; 2016; 2018; Smolcic et al. 2017; Delvecchio et al. 2017; Sabater et al. 2019). A similar trend (even though with a much wider mass distribution --  cfr Figure 9) is also present in the behavior of X-ray-selected sources and is  due to the combination of observational bias effects with the fact that AGN are more common in more massive galaxies (e.g. Silverman et al. 2009; Aird et al. 2012; Aird et al. 2013; Bongiorno et al. 2012; Suh et al. 2017;  Georgakakis et al. 2017; Yang et al. 2018)\footnote{We stress that Figure 8 only represents the fraction of sources active in the X-ray or radio bands which are found to host an AGN at their center as a function of the stellar mass of the hosts, and 'per se' does not provide any information on the general underlying population of active+inactive galaxies, at variance with the works of e.g. Brusa et al. (2009),  Bongiorno et al. (2016),  Suh et al. (2017) and Georgakakis et al. (2017) and Mezcua et al. (2018)}. 

However, amongst X-ray emitters in the CDFS (which we remind is the deepest of the two fields) we find a non negligible tail of low-mass, $M_*\simlt 10^{9} M_\odot$, AGN whose relative contribution becomes more and more prominent as the mass of the host galaxy is lowered. 

We further investigated this result by inspecting such $M_*\simlt 10^{9} M_\odot$ sources one by one, in order to make sure there were no issues in the photometry and therefore in the mass estimates derived from them. We also looked for the mass values obtained by the CANDLES collaboration (Santini et al. 2009) by varying assumptions for the star-formation history, initial mass function, nebular emission contribution and other parameters. All these tests confirmed the low masses obtained for these sources (37 AGN, out of which 12 with $M_*<10^{8} M_\odot$, all but two found in the nearby, $z<1$, universe due to the completeness mass limits of the parent VANDELS catalogues - cfr McLure et al. 2018). 


Another possibility is that the X-ray signal produced by these low-mass sources is due to the activity of X-ray binaries within the host galaxy and that these objects were therefore mistakenly classified as AGN by Luo et al. (2017). However, the energetics associated to these events would have to be extremely high, all above the range of values  $10^{35}$ erg sec$^{-1}$ $\simlt  L_X \simlt 10^{39}$ erg sec$^{-1}$ expected for the local, $z\simlt 1$, population of X-ray binaries associated with galaxies of such low masses (e.g. Fragos et al. 2013).  Also, the star-formation rates of these sources are compatible with an X-ray 
emission dominated by AGN mechanisms (Mineo et al. 2014).

The last possibility is that these low-mass AGN are the result of mismatches, since the chances of spurious associations increase when we restrict the VANDELS catalogue to galaxies of low mass. In order to quantify this effect, we have compared the redshifts of the sample of $M_*<10^{9} M_\odot$ X-ray AGN  as obtained from VANDELS with those assigned to these sources by the Luo et al. (2017) work. This is because, although the fact that a source with different redshifts in the two catalogues does not automatically imply a mismatch, indeed an object which has similar redshifts from our work and from that of Luo et al. (2017) who use different matching techniques can be considered as correctly associated to its counterpart. It turns out that 6 out of 12 $M_*<10^{8} M_\odot$ AGN and 19 out of 25 $10^8 M_\odot<M_*<10^{9} M_\odot$ AGN in our sample have redshifts in good agreement (differences within 14\%) with those from the Luo et al. (2017) catalogue. We can therefore conclude that at least 50\% of the very low-mass,  $M_*<10^{8} M_\odot$, AGN and at least 76\% of the $10^8 M_\odot<M_*<10^{9} M_\odot$ AGN in our sample are indeed true associations.  

The above discussion implies that the trend shown in the left-hand panel of Figure  8 is indeed a true feature of our sample which points to a non-negligible presence of  X-ray emitting AGN within low-mass galaxies. We note that our result confirms that of Mezcua et al. (2018) obtained on the COSMOS field on the existence of X-ray AGN within the population of dwarf galaxies  ($10^7$ M$_\odot<M_*<10^{9}$ M$_\odot$), and possibly extends it  down to the mass range $10^6 M_\odot\simlt M_*\simlt 10^7 M_\odot$, where we find 3 AGN. We further note that our finding seems to be connected with the AGN luminosity in the X-ray band. Indeed, we only observe AGN in low-mass galaxies in the CDFS, while none is present in galaxies below $M_*=10^8 M_\odot$ either in the UDS or in the CDFS for AGN luminosities $L_X \ge 10^{43}$ erg sec$^{-1}$. 



The mass distributions of the samples of AGN as obtained in the CDFS and UDS fields are presented in Figure 9. The top panels show the cases for all AGN emitting in the X-ray (dotted histograms), for only those with X-ray luminosities above the completeness limits of the two surveys (short-long dashed histograms), for radio-active AGN (solid histograms) and for those which emit at both radio and X-ray wavelengths (dashed histograms). The bottom panels instead show the ratios between the above quantities. More quantitative information can be found in Tables 5 and 6. As it was in the case of the redshift distributions, our data indicates that the mass distributions of the hosts of AGN active in the radio or X-ray bands are quite different. Indeed, as previously discussed, 
the mass distribution of X-ray emitting AGN, especially in the case which also includes luminosities $L_X\simlt 10^{42.5}$ erg sec$^{-1}$, extends to rather low masses. Such a low-mass tail does not show in the mass distribution of radio-emitting AGN which only start appearing within  host galaxies of masses $M_*\simgt 10^{10} M_\odot$. Furthermore and  regardless of $L_X$, while the mass distribution of X-ray AGN has a peak in the range $10^{10} M_\odot\simlt M_*\simlt 10^{11} M_\odot$ and then sensibly decreases at larger masses, that of radio-active AGN is almost constant between $10^{10} M_\odot$ and $10^{12} M_\odot$.

No decline is either seen in the mass distribution of those AGN which are simultaneously active at both radio and X-ray wavelengths as their number keeps monotonically rising with host mass up to the largest stellar masses probed by our analysis (dashed histograms in the top panels of Figure 9). The net effect of these combined trends is shown in the bottom panels of Figure 9: the relative number  of AGN which are active at both X-ray and radio wavelengths is a sharply increasing function of the stellar mass of their hosts. This happens in comparison with both the parent population of X-ray AGN (dashed histograms) and with that of radio AGN (solid histograms), even though the effect is visually more striking in this latter case.

Investigations of Tables 5 and 6 show that the combined probability for an X-ray emitting AGN to also be active at radio wavelengths increases from $\sim 1$\% to $\sim 13$\% when the stellar mass its host increases from $\sim 10^{10.5} M_\odot$ to $\sim 10^{11.5} M_\odot$. Within the same mass range, the chances for a radio-active AGN to also emit in the X-ray  go from $\sim 15$\% to $\sim 45$\%. The rise is more striking in the case of CDFS (increment from $\sim 19$\% to $\sim 78$\%, cfr left-hand panel of Figure 9) than for the UDS (increment from $\sim 13$\% to $\sim 23$\%, cfr right-hand panel of Figure 9). This is most likely due to the different depths of the two X-ray surveys.


A summary of the previous findings is presented in Figure 10. Here we draw two pie charts representing the entire AGN population obtained in the combined CDFS+UDS fields by adding up all AGN which are active at radio or X-ray wavelengths. The yellow portions indicate those AGN which only emit in the X-ray, the purple portions those which only emit in the radio, while the green portions represent the percentages of AGN which are active in both bands. If we consider the whole mass range probed by the host galaxies (left-hand plot), we find that the overwhelming majority of the 
AGN population ($\sim 90$\%) is only active in the X-ray. About 7.5\% is only active in the radio and only $\sim 2.5$\% shows activity at both frequencies.

However, the situation dramatically changes when one restricts to AGN hosted by high-mass ($M_*\ge 10^{11} M_\odot$) galaxies (right-hand plot and associated values). Indeed, in this case we find that
about 14\% of the whole AGN population is made of those which are only active in the radio band, and about 11\% are those which emit at both frequencies. We then have that the contribution of these latter sources to the whole AGN population  increases by almost a factor 
$5$ when moving from the all-mass to the high-mass regime, and cannot be explained by simply advocating a more prominent presence of radio-active AGN within high-mass galaxies. In fact, as clearly shown by the two pie charts in Figure 10 (but also see Figure 9),  the relative contribution of radio-emitting AGN  increases by less than a factor $2$ when moving from the all-mass AGN population to the sub-population of AGN harboured within hosts of large stellar mass.  

Our analysis therefore indicates that {\it the mass of a galaxy host of an AGN plays a crucial role in determining the AGN level of activity (expressed in terms of luminosity)} at the various wavelengths. 
Noticeably, this not only holds for radio or X-ray emission alone, but even more so {\it is linked to simultaneous AGN activity in the various bands of the electromagnetic spectrum}.

\begin{figure*}
\includegraphics[scale=0.4]{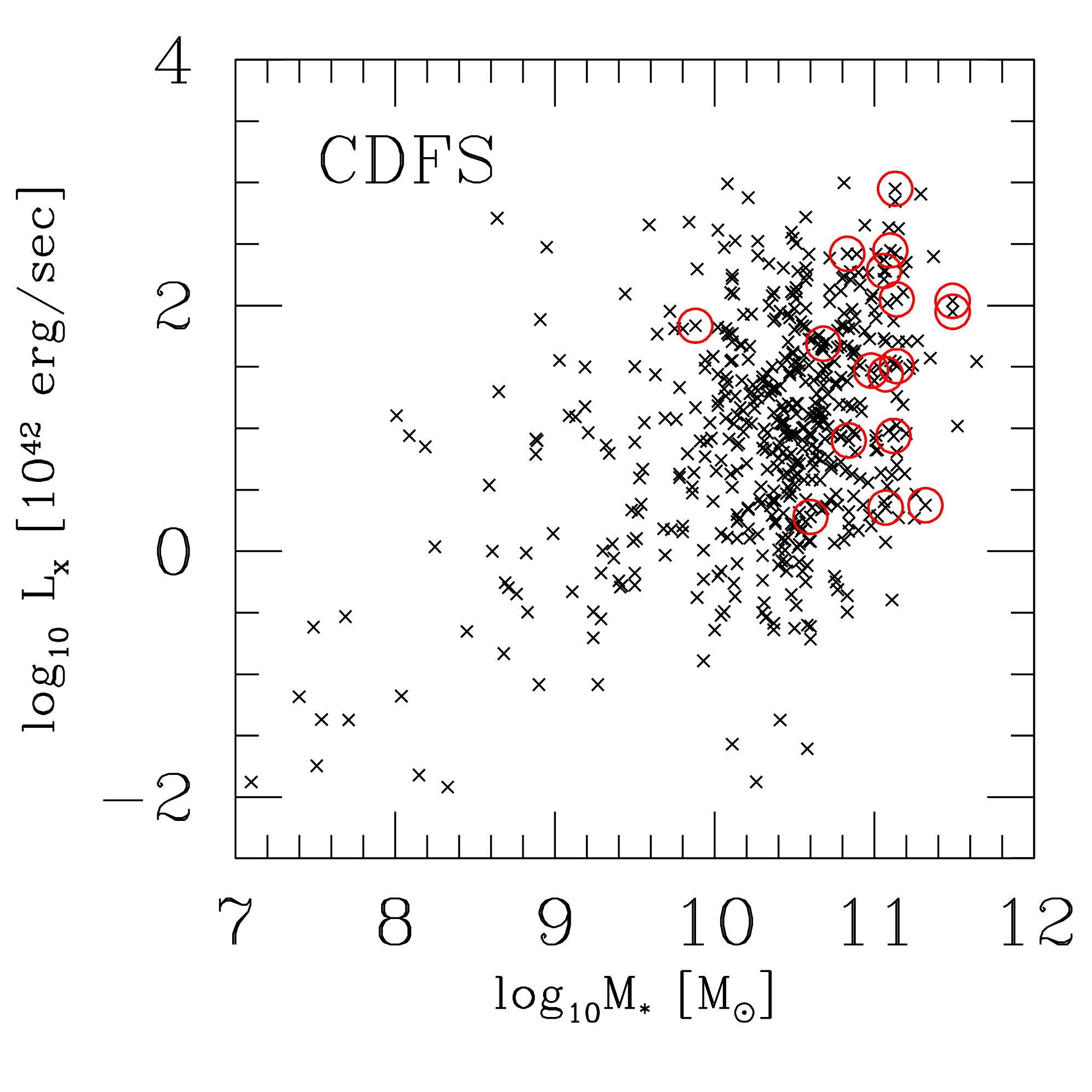}
\includegraphics[scale=0.4]{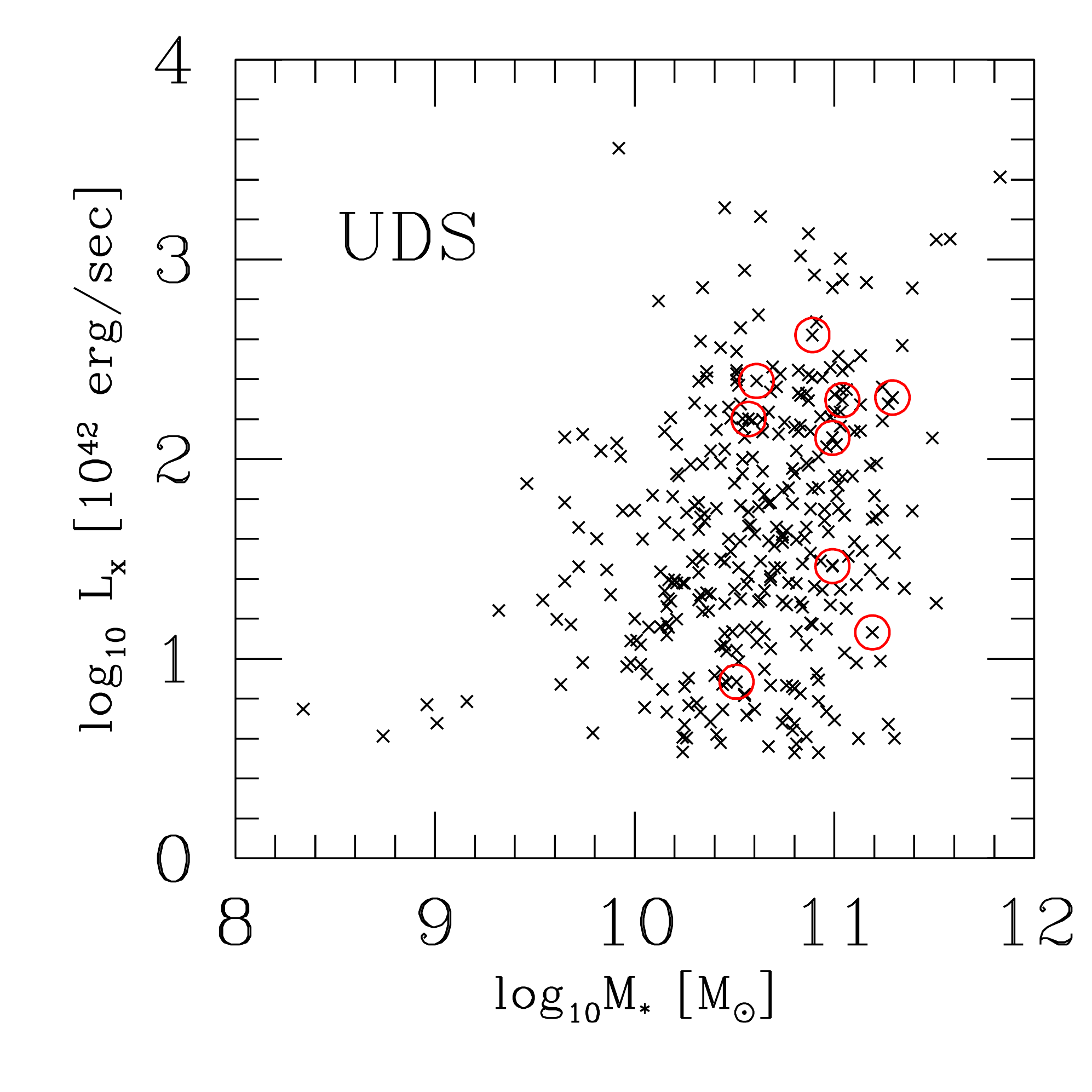}
\caption{Distributions of X-ray luminosities and stellar masses for X-ray-selected AGN on the CDFS (left-hand panel) and UDS (right-hand panel). Open circles indicate those X-ray-emitting AGN which have an AGN counterpart at radio wavelengths. Note the different scales for $L_X$ in the two panels.}
\end{figure*}

 \begin{figure*}
\includegraphics[scale=0.4]{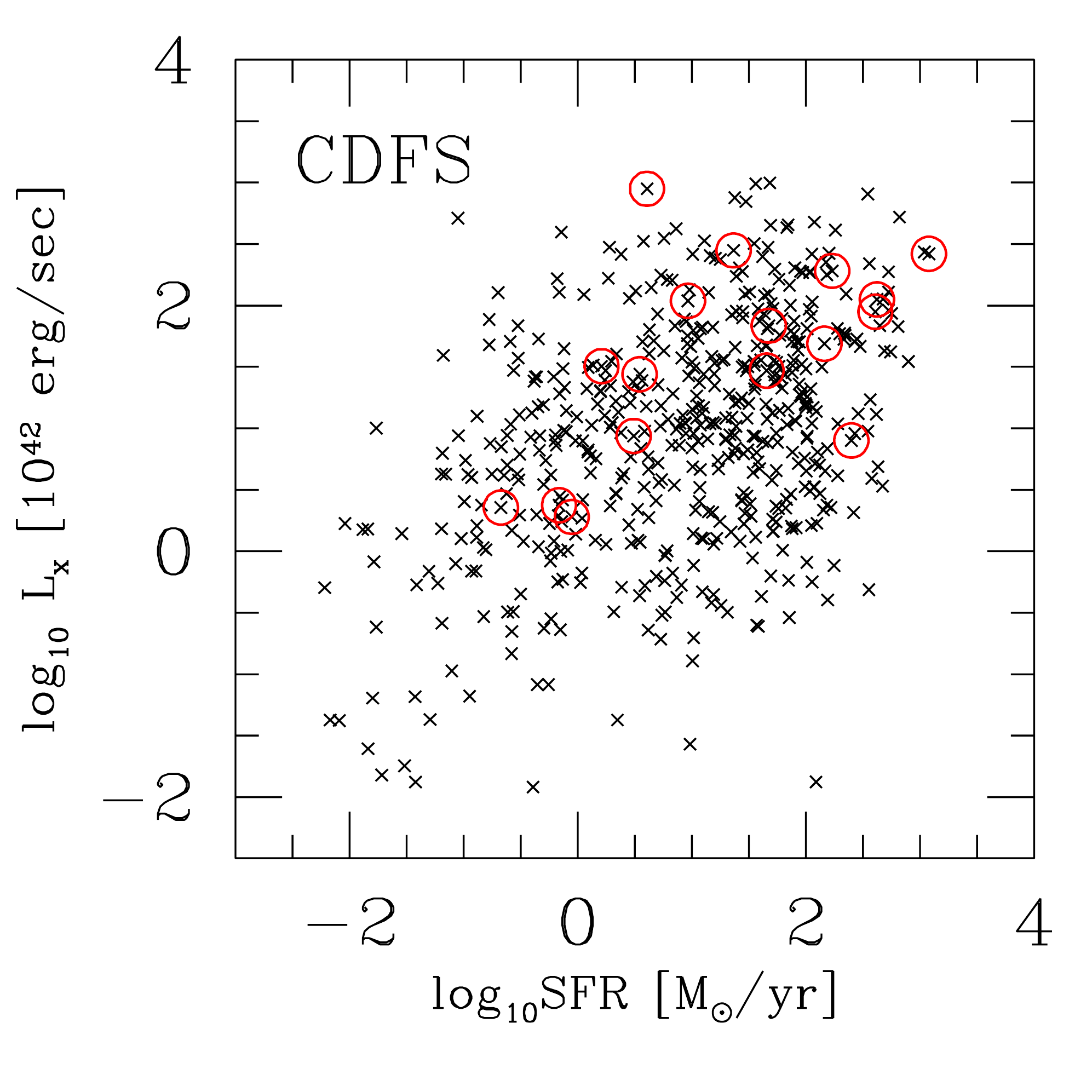}
\includegraphics[scale=0.4]{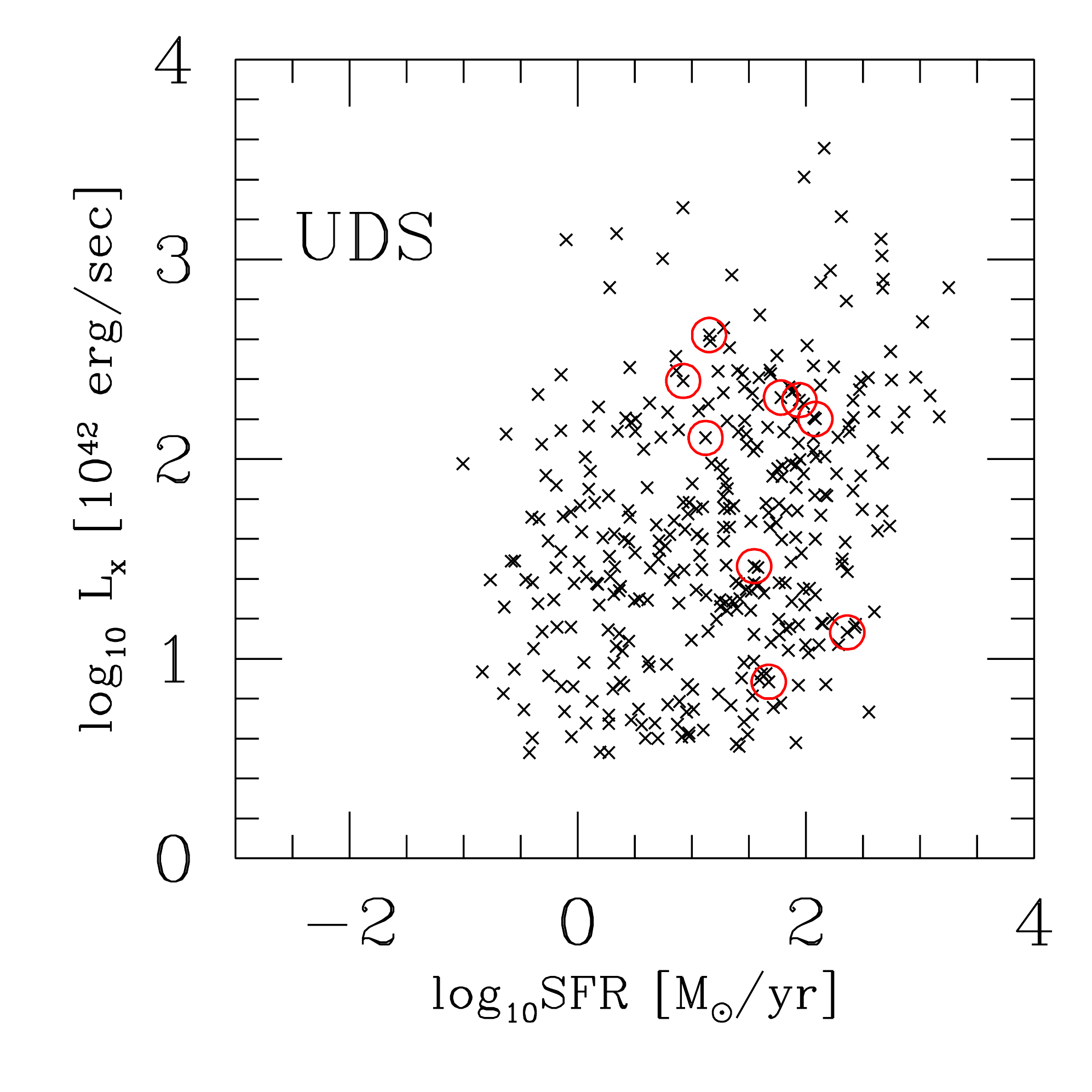}
\caption{Distributions of  X-ray luminosities and star-formation rates for X-ray-selected AGN on the CDFS (left-hand panel) and UDS (right-hand panel). Open circles indicate those X-ray-emitting AGN which have an AGN counterpart at radio wavelengths. Note the different scales for $L_X$ in the two panels.}
\end{figure*}	

 \begin{figure*}
\includegraphics[scale=0.4]{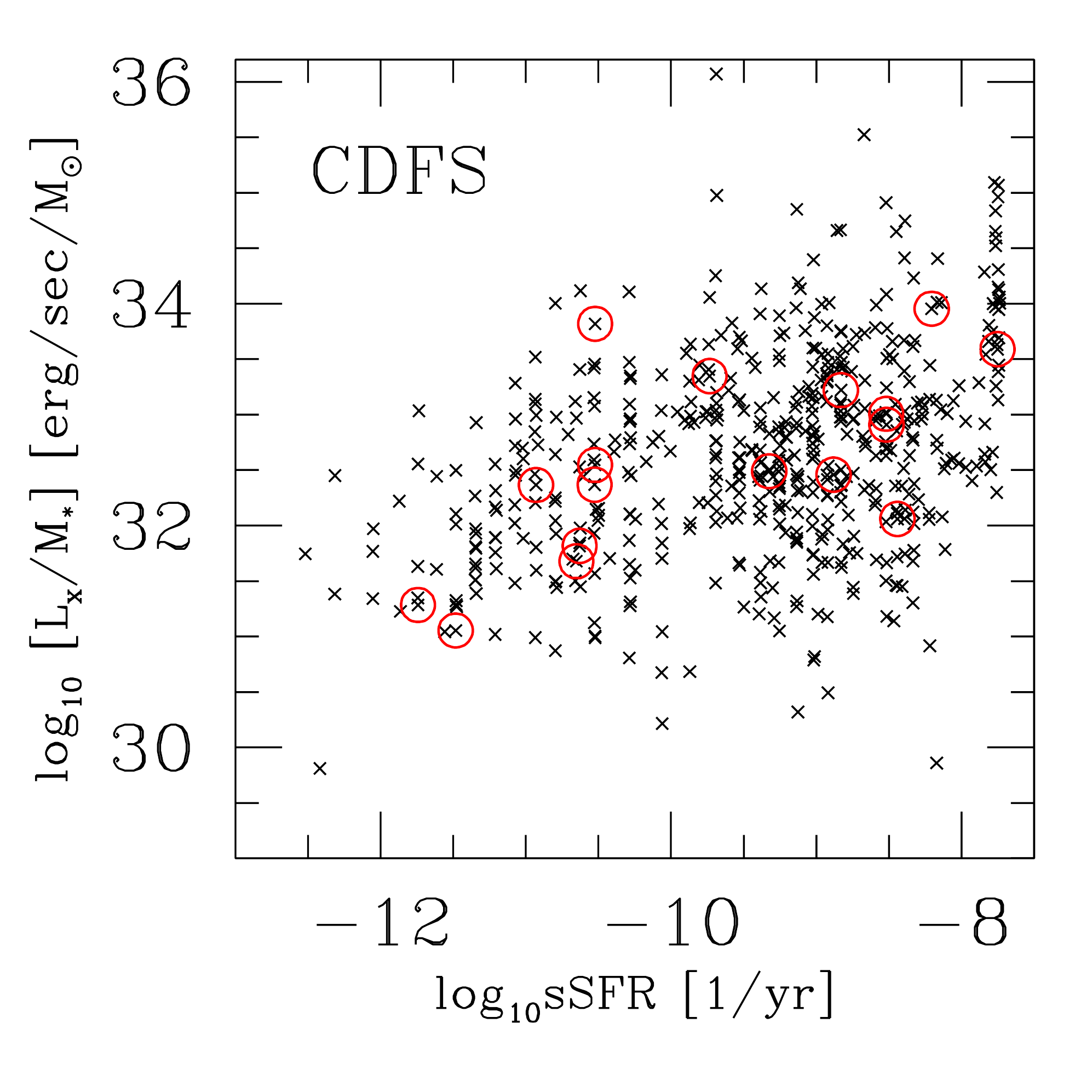}
\includegraphics[scale=0.4]{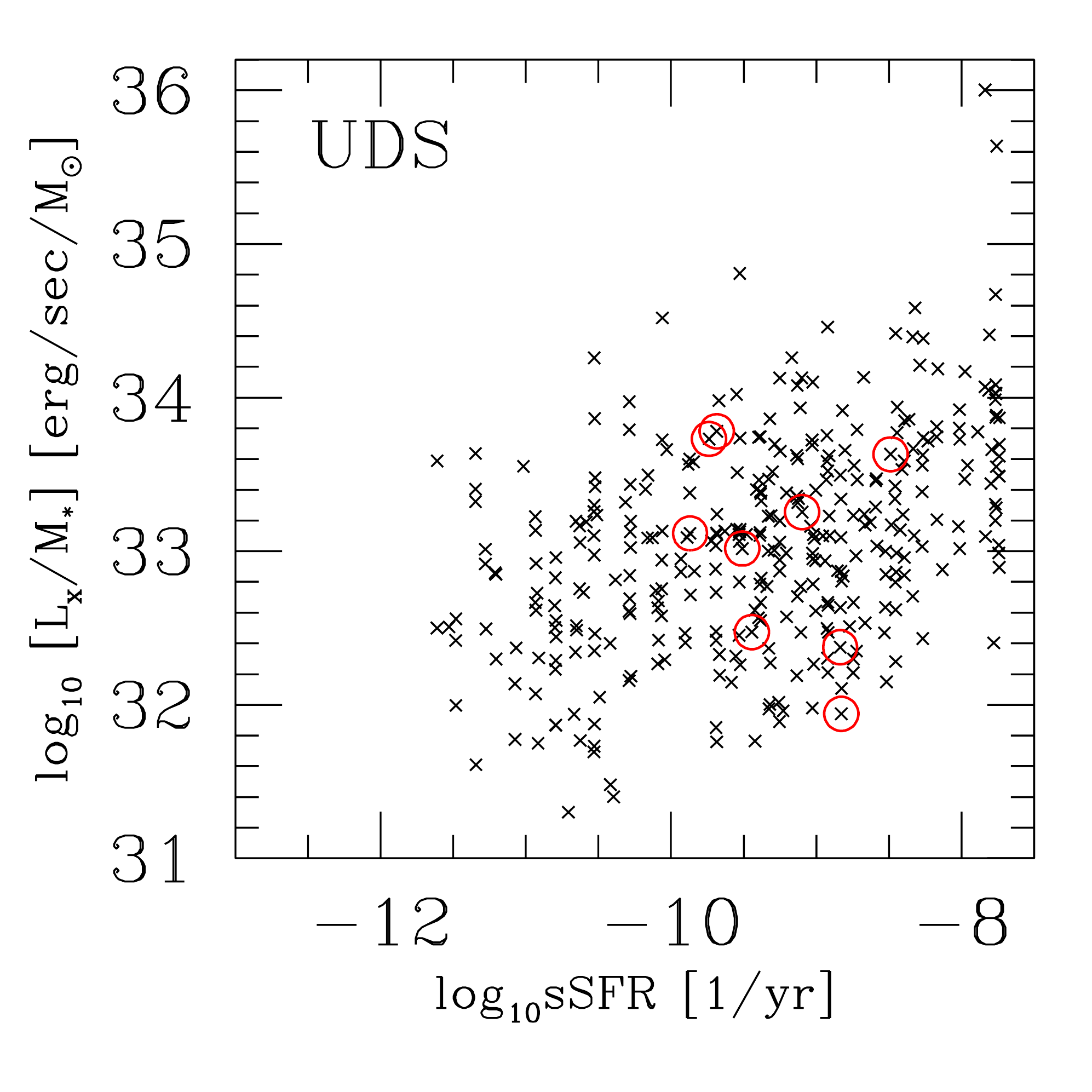}
\caption{Distributions of  X-ray luminosities normalized by the mass of the hosting galaxy vs specific star-formation rates for X-ray-selected AGN on the CDFS (left-hand panel) and UDS (right-hand panel). Open circles indicate those X-ray-emitting AGN which have an AGN counterpart at radio wavelengths. Note the different scales for $L_X/M_*$ in the two panels.}
\end{figure*}

 \begin{figure*}
\includegraphics[scale=0.4]{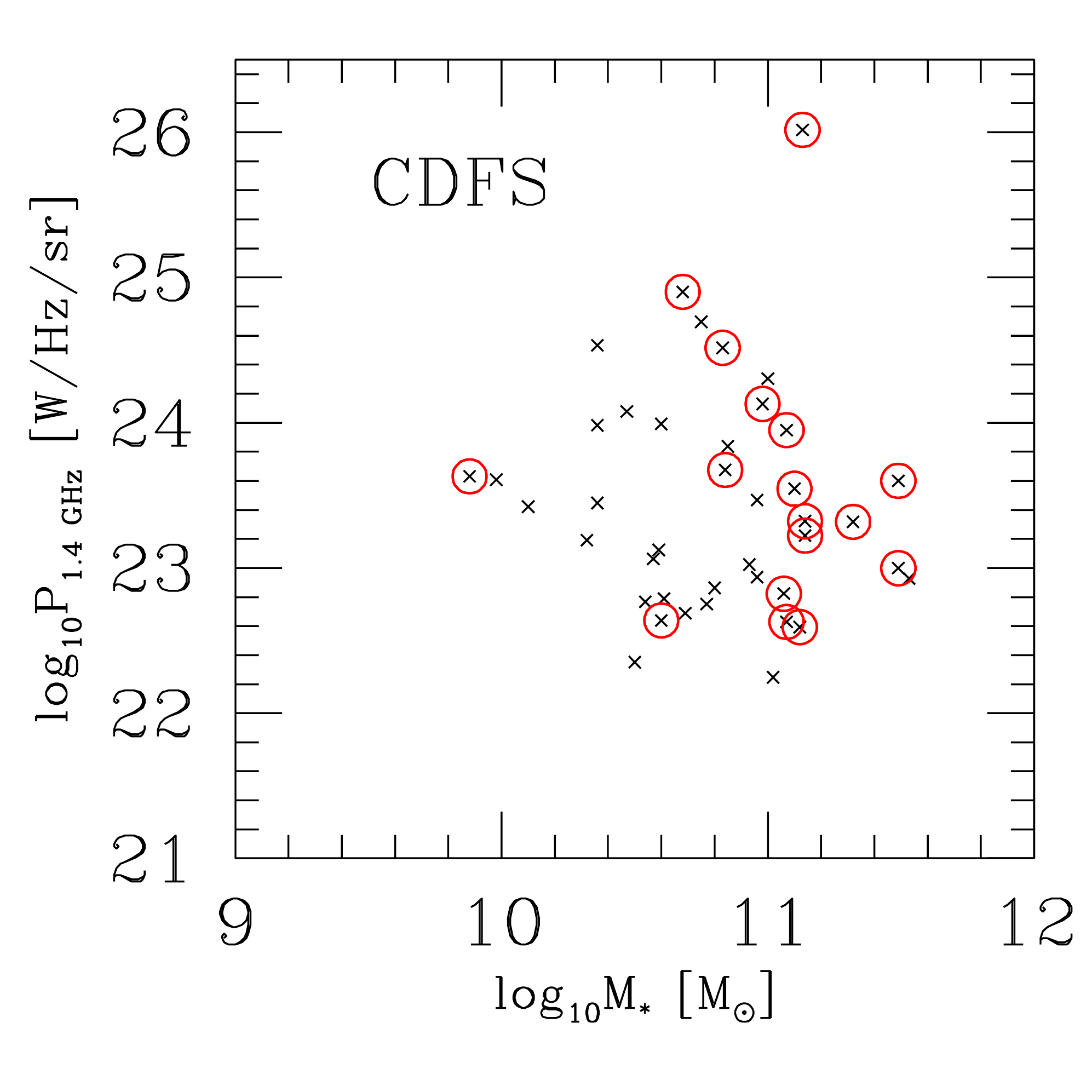}
\includegraphics[scale=0.4]{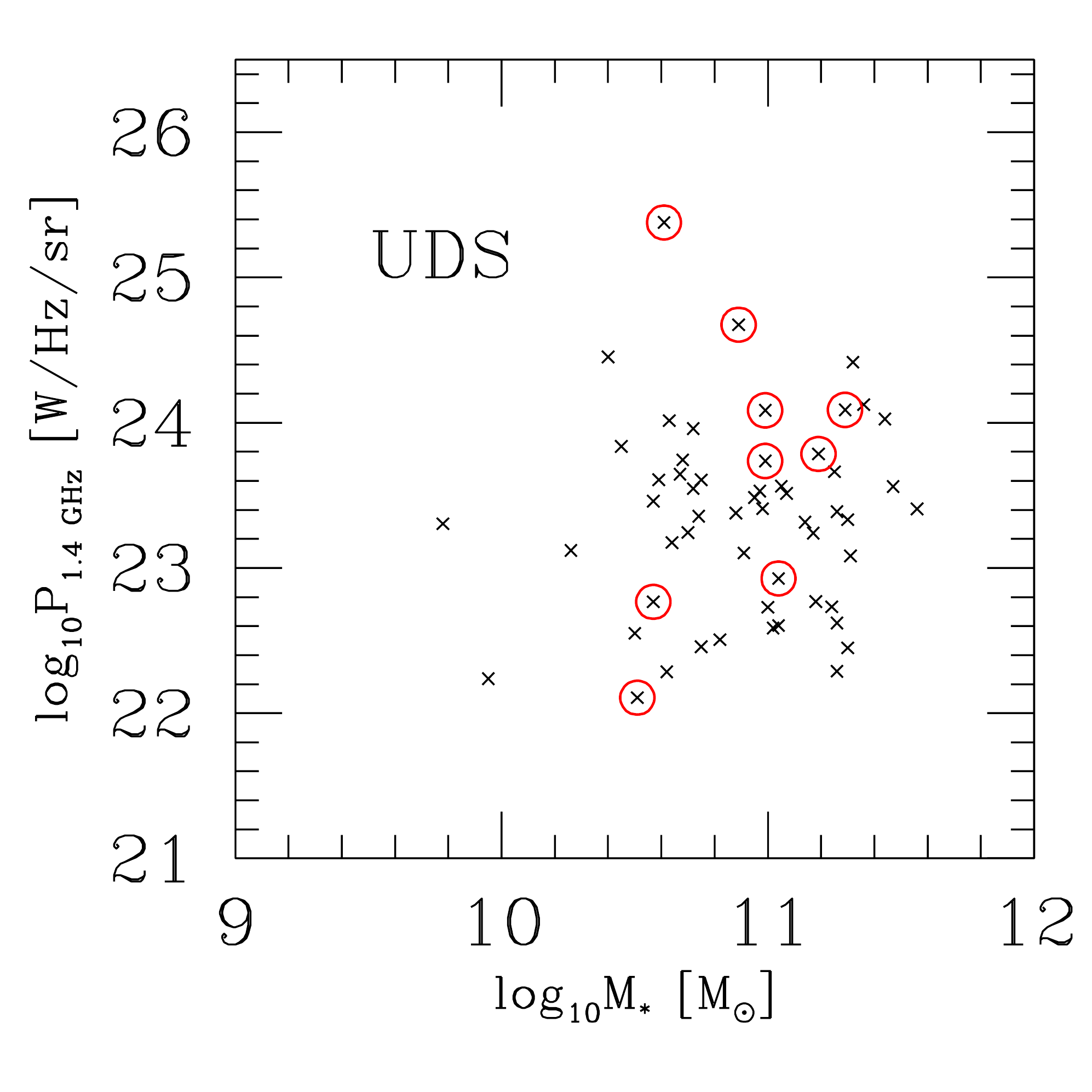}
\caption{Distributions of radio luminosities and masses for radio-selected AGN on the CDFS (left-hand panel) and UDS (right-hand panel). Open circles indicate those radio-emitting AGN which have an AGN counterpart at X-ray wavelengths.}
\end{figure*}

 \begin{figure*}
\includegraphics[scale=0.4]{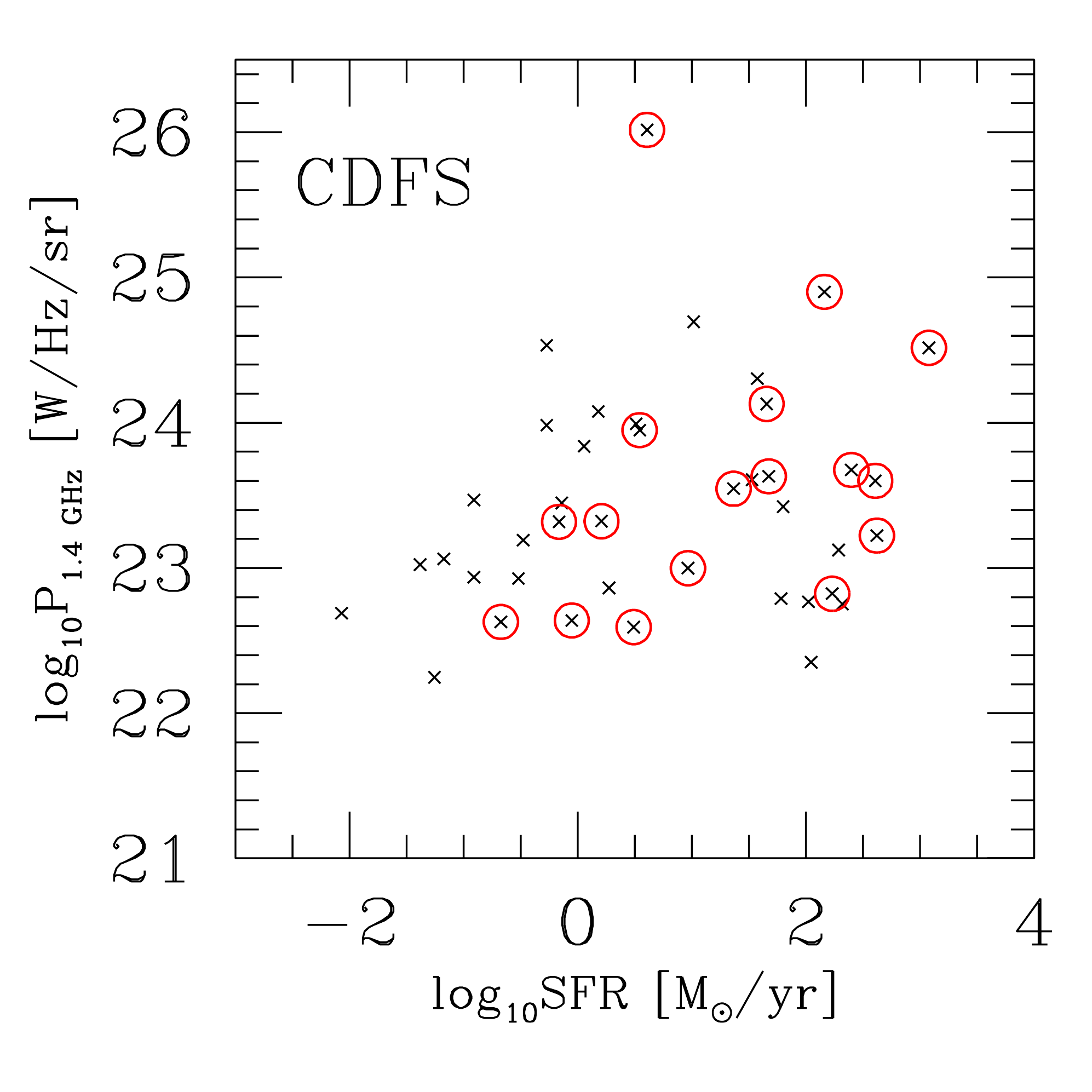}
\includegraphics[scale=0.4]{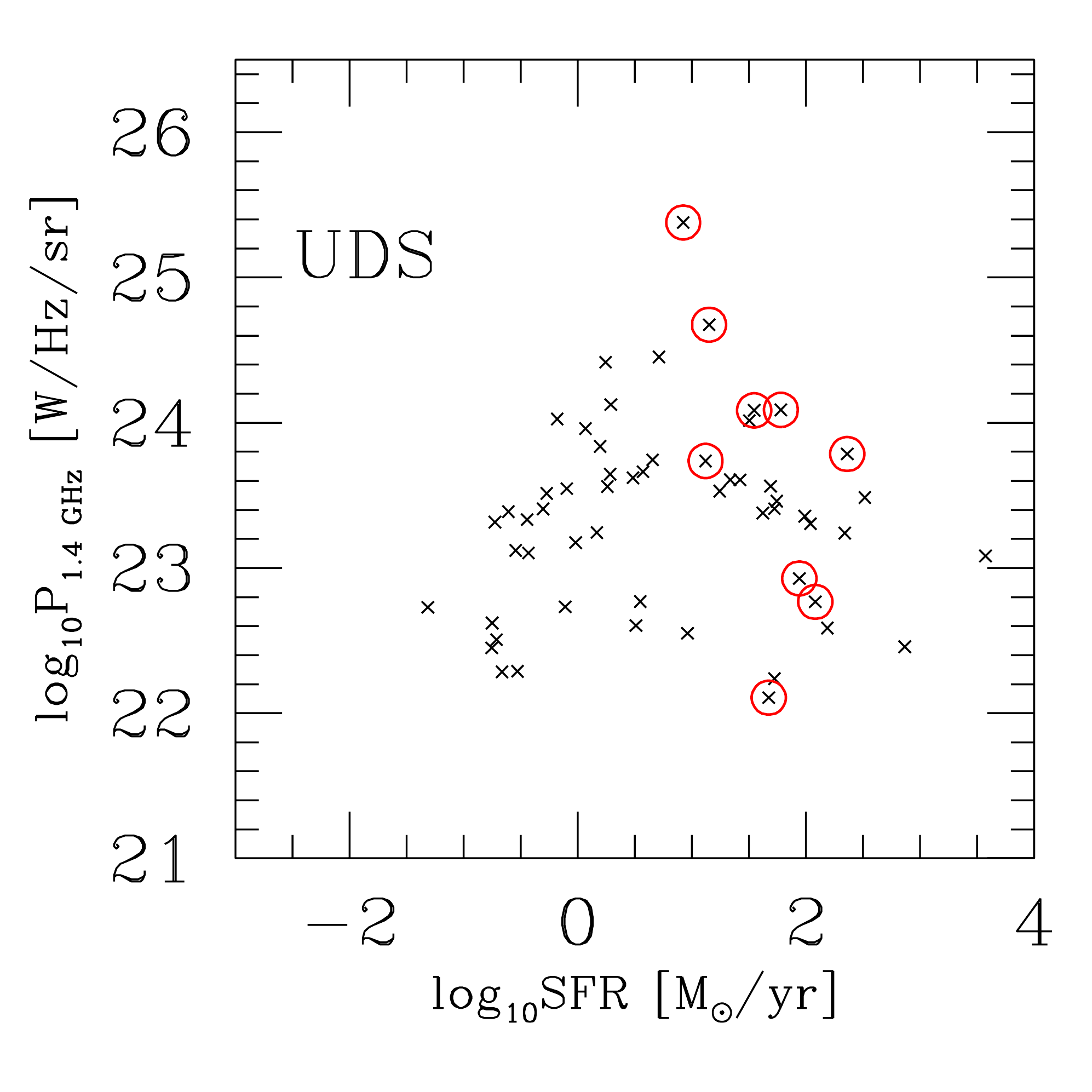}
\caption{Distributions of  radio luminosities and star-formation rates for radio-selected AGN on the CDFS (left-hand panel) and UDS (right-hand panel). Open circles indicate those radio-emitting AGN which have an AGN counterpart at X-ray wavelengths.}
\end{figure*}	
 
\begin{figure*}
\includegraphics[scale=0.4]{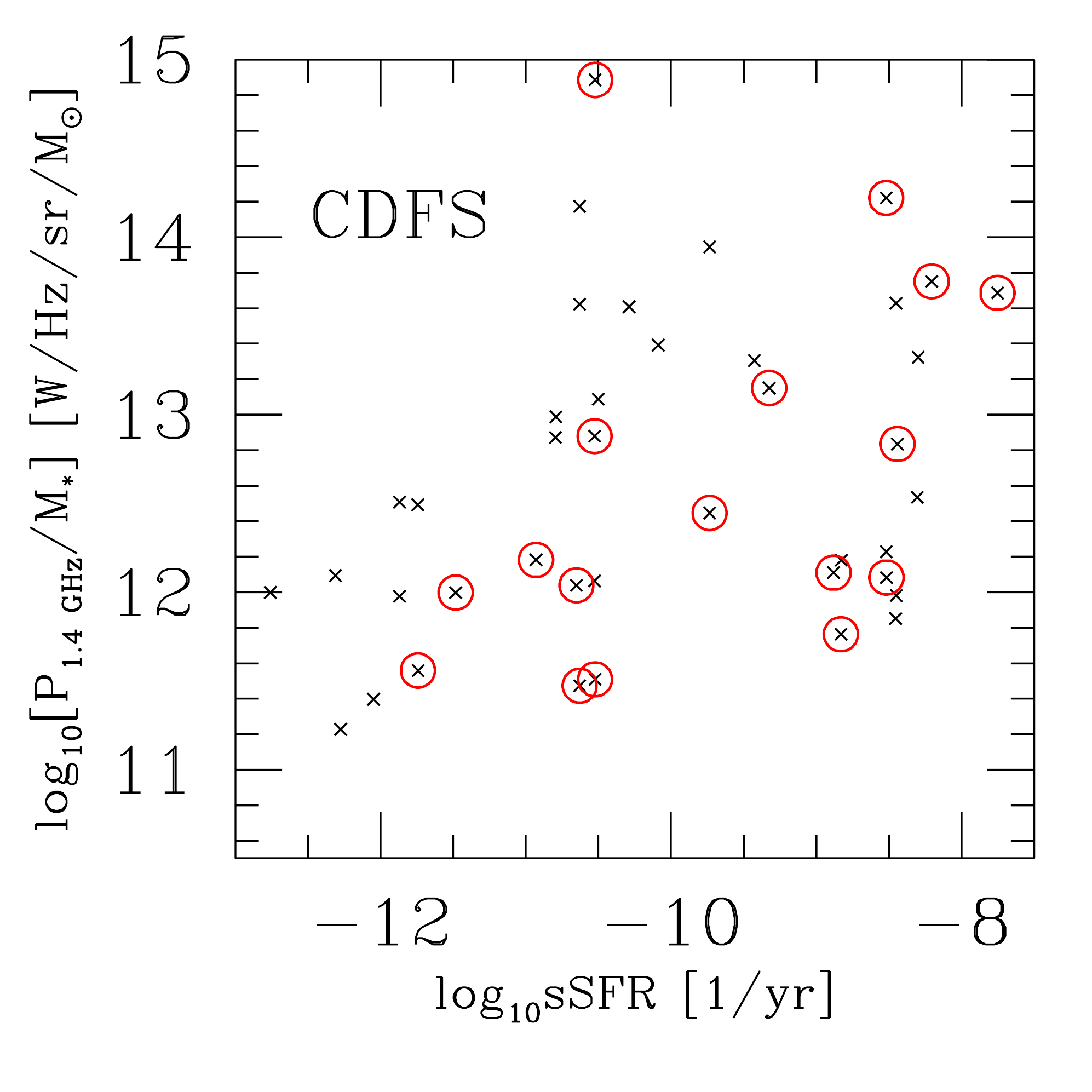}
\includegraphics[scale=0.4]{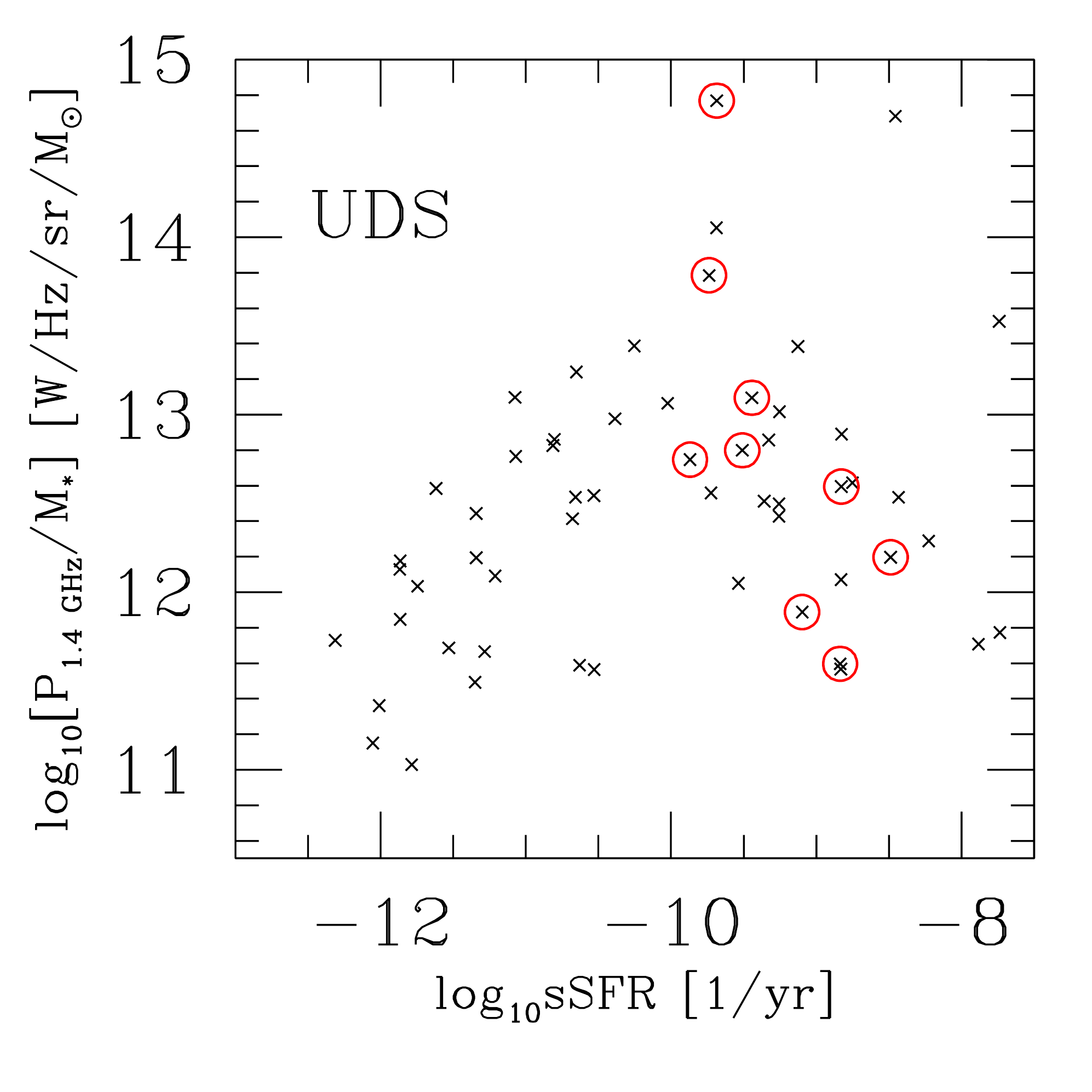}
\caption{Distributions of  radio luminosities normalized by the mass of the hosting galaxy vs specific star-formation rates for radio-selected AGN on the CDFS (left-hand panel) and UDS (right-hand panel). Open circles indicate those radio-emitting AGN which have an AGN counterpart at X-ray wavelengths.}
\end{figure*}	

\section{is mass the only relevant parameter?}
In the previous Section we have shown how the mass of a galaxy host of an AGN is crucially linked to AGN activity at the various wavelengths.  This is especially true in the case of simultaneous emission in the X-ray and radio bands as our data indicate that about 11\% of all AGN residing within galaxies of masses $\ge 10^{11} M_\odot$ ($\sim 45$\% if we restrict to the parent population of radio-active AGN, percentage which increases to $\sim 80$\% if we only consider the CDFS which samples fainter X-ray luminosities) emit at both frequencies.\\ 
But is galaxy mass the only relevant physical quantity to favour combined X-ray+radio AGN emission?

In order to answer this question, we have investigated the behavior of a number of properties which characterize the sources under exam, both those selected at 1.4 GHz and those selected in the X-ray. As a first step, in the case of AGN emitting at X-ray wavelengths, we show the distribution of X-ray luminosities  and masses (Figure 11). As it was for the previous Figures, the left-hand panel refers to CDFS, while the right-hand panel is for UDS. Note the different scales for $L_X$ in the two panels. Crosses refer to the whole X-ray AGN population, open circles to those AGN which also emit in the radio.

Apart from the remarkable preference for AGN simultaneously emitting at both radio and X-ray frequencies to reside within high-mass galaxies already discussed in \S 5, Figure 11 hints to a 
luminosity dependence whereby, especially in the CDFS case (left-hand panel), radio-AGN emission is uniquely associated to X-ray-bright, $L_X>10^{42}$ erg sec$^{-1}$, sources.  
However, we remind that $L_X=10^{42}$ erg sec$^{-1}$ is the limit for completeness of the Luo et al. (2017) survey. This implies that we are likely missing X-ray AGN with luminosities fainter than the above value which could  in principle have a radio-AGN counterpart. Similarly, the paucity of X-ray AGN associated with radio emission below $L_X=10^{43}$ erg sec$^{-1}$ observed in the UDS (right-hand panel) can be easily explained as due to the completeness limit of the X-ray AGN sample within this field (cfr \S 4.2). 
We can then conclude that, on the basis of our available data, there does not seem to be any compelling evidence for X-ray luminosity to be amongst the main factors which determine simultaneous X-ray and radio emission from an AGN.
Also, as illustrated in Figure 12, there only seems to be a very mild dependence on the star-formation rate of the hosts, only visible in the UDS. The same results are also obtained if we factor out the mass dependence in Figure 12 and consider the distribution of specific luminosities $L_X/M_*$ versus specific star-formation rates (sSFR), defined as SFR$/M_*$ (cfr Figure 13).
However, we remind once again that the UDS only probes bright X-ray AGN which are preferentially associated with a higher level of star-formation activity within their hosts (e.g. Santini et al. 2012;  Rosario et al. 2015, even though the subject is still under debate, cfr Brusa et al. 2009; Del Moro et al. 2016). This suggests that the trends shown in the right-hand panels of Figures 12 and 13 are most likely due to the relative shallowness of the X-ray survey performed on the UDS. Indeed, in the CDFS which probes much fainter luminosities, we do not see any dependence on the star-formation rates of the AGN hosts (left-hand panels of Figure 12 and 13). As a further test aimed at removing possible luminosity dependencies, in the CDFS we have only considered the 278 AGN with X-ray luminosities above $L_X=10^{43}$ erg sec$^{-1}$ (12 with a radio-AGN counterpart) which, we remind, is the limit for completeness in the UDS. In this case, the distributions of SFRs and SSFRs as a function of both X-ray luminosity and radio luminosity  show a deficit of AGN simultaneously emitting in the radio and X-ray bands for low SFRs ($\simlt 10$ M$_\odot$/yr) and SSFRs ($\simlt 10^{-10}$ 1/yr) with respect to the original CDFS case (4 sources instead of 8), and a total lack of such objects below SFR $\sim 1$ M$_\odot$/yr and SSFR $\sim 10^{-11}$ 1/yr, in much better agreement with the UDS results.
We can then also conclude that, on the basis of our available data, there does not seem to be any compelling evidence for the star-formation rate to be amongst the main factors which determine simultaneous X-ray and radio emission from an AGN.

\begin{figure}
\includegraphics[scale=0.4]{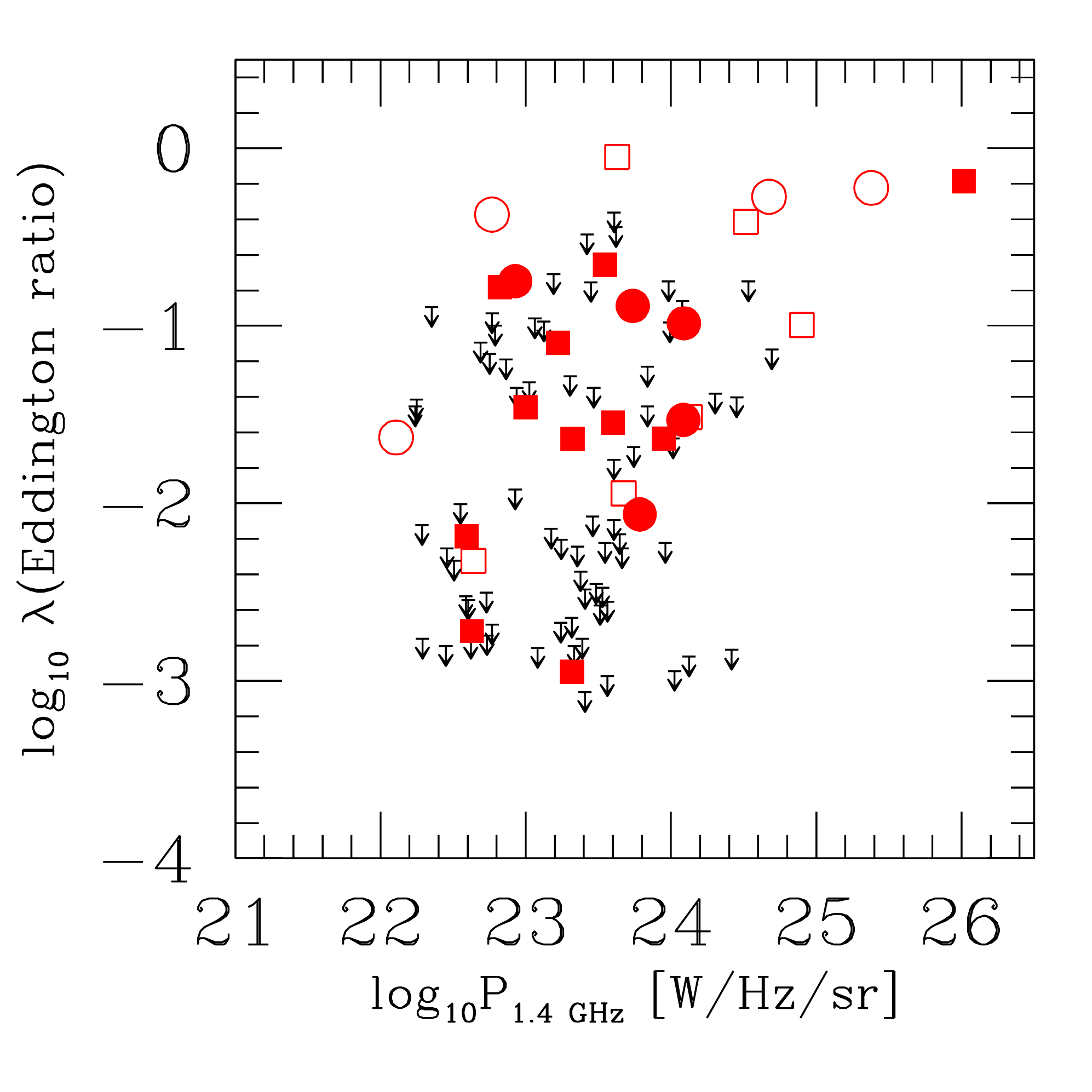}
\caption{Distribution of  Eddington ratios vs radio luminosities for those AGN which simultaneously emit in both X-ray and radio bands. AGN in the CDFS are represented by squares while those in the UDS by circles. Full symbols indicate sources associated with galaxies of masses $M_*\ge 10^{11} M_\odot$. Upper limits correspond to radio-AGN in both the CDFS and UDS fields without a detected X-ray AGN counterpart.}
\end{figure}

\begin{figure}
\includegraphics[scale=0.4]{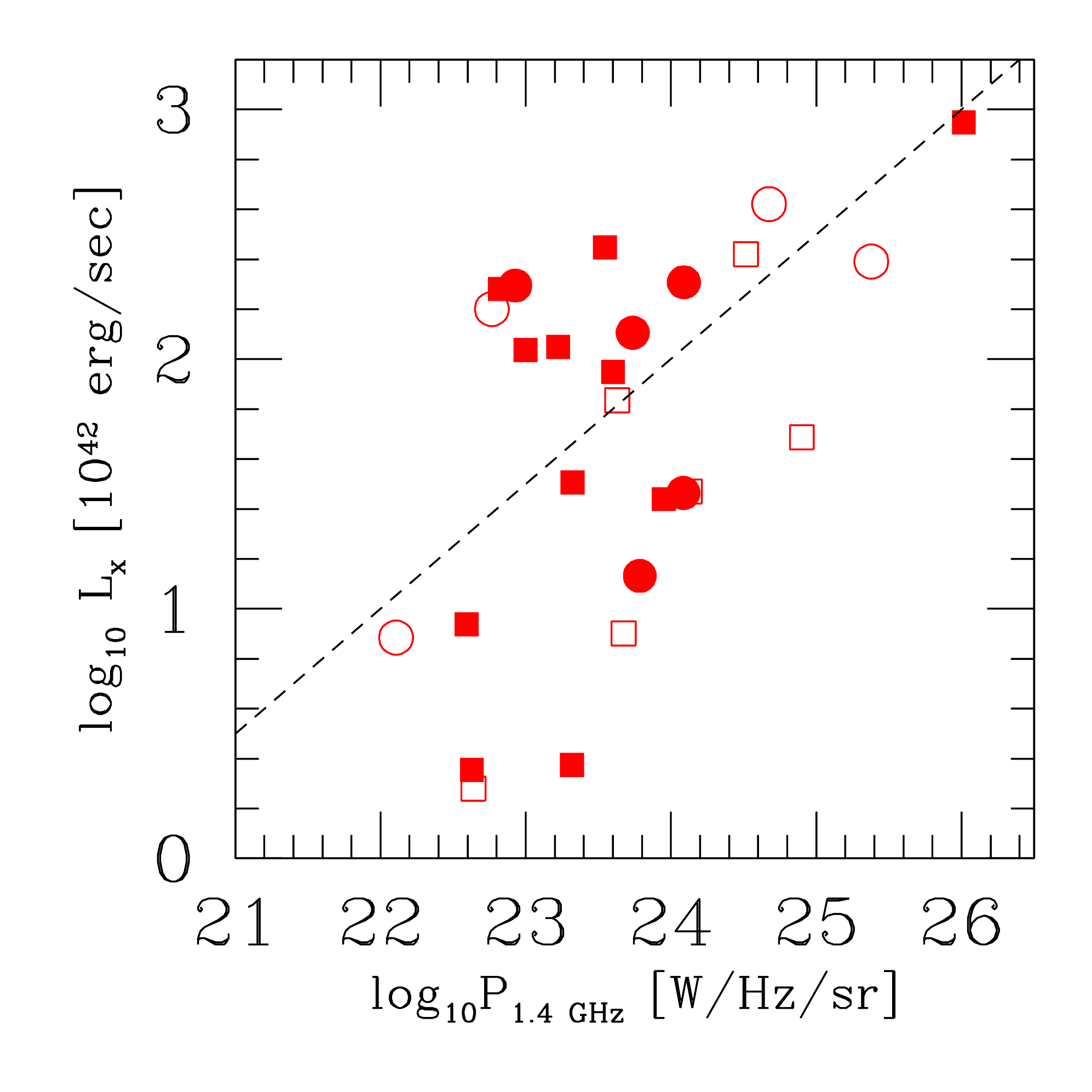}
\caption{Distributions of  radio and X-ray luminosities  for those AGN which simultaneously emit in both bands as found in the CDFS (squares) and UDS (circles). Full symbols indicate sources associated with galaxies of masses $M_*\ge 10^{11} M_\odot$. The dashed line indicates the best linear fit to the data (cfr text for further information).}
\end{figure}	

A similar analysis can be performed on the properties of radio-selected AGN and of their hosts. This is done in Figures 14, 15 and 16 which respectively refer to the distributions of radio luminosity as a function of stellar mass of the hosts, radio luminosity as a function of star-formation rate and specific radio luminosity $P_{\rm 1.4 GHz}/M_*$ as a function of specific star-formation rate defined as before. Here crosses are for all radio-selected AGN, while circles represent those sources which are also active in the X-ray. Similarly to what found in the case of X-ray selected AGN, we observe a clear evidence for those which also emit at X-ray wavelengths to be associated with galaxies of large, $M_*>10^{10.5} M_\odot$, stellar mass content and a mild trend to appear within galaxies with relatively high star-formation rates. However, once again this is only true for sources coming from the UDS and, as explained in the previous paragraph, is most likely due to selection effects. At the same time, it is clear from Figures 14 and 15 that there is no dependence on radio luminosity, as those sources which simultaneously emit at both frequencies span all possible ranges of $P_{\rm 1.4 GHz}$. In other words, {\it the chance for an AGN to be simultaneously active at both radio and X-ray wavelengths does not depend on either its X-ray or radio luminosity and neither it depends on the star-formation activity within the host galaxy}. 


Another physical quantity which is relevant when considering AGN populations is the Eddington ratio, which measures the growth rate of a black hole relative to its maximum capacity. Under appropriate assumptions, the Eddington ratio can be expressed in terms of the AGN X-ray luminosity and of the stellar mass of its host galaxy as follows (cfr Georgakakis et al. 2017):
\begin{equation} 
\lambda=\frac{25 \cdot L_X({\rm 2-10 KeV)}}{1.26\cdot 10^{38}\cdot 0.002 M_*}
\end{equation}

Equation (3) was then applied to our sample of radio-AGN and estimates for the Eddington ratio obtained for both X-ray detections and non detections are plotted in Figure 17.  Circles correspond to UDS sources, squares to CDFS sources, while upper limits at the source redshifts have been derived according to Figures 2 and 4.

Figure 17 highlights a preference for AGN detected at both radio and X-ray wavelengths and associated to relatively low-mass, $M_*< 10^{11} M_\odot$, galaxies to present high Eddington ratios: indeed, six of such sources out of ten (shown by the open symbols) exhibit values  $\lambda>-1$. Five of them have $\lambda>-0.4$. There might also be a hint for very bright, $P_{1.4 \rm GHz}\simgt 10^{24.8}$ [W/Hz/sr], radio-AGN to be always associated with X-ray emission and with high values of $\lambda$, but our statistics is currently too scanty to allow for firm conclusions to be drawn. No other trend can be identified: we do not observe any obvious dependence of $\lambda$ on radio luminosity, not even in the distribution of Eddington ratios obtained for radio-AGN without an X-ray AGN counterpart.
We can then conclude that {\it simultaneous radio and X-ray emission of AGN origin is also not related to the accreting activity of the black hole responsible for the AGN signal.} 

Furthermore, as shown in Figure 18, once an AGN is active in both bands, there does not seem to be any strong relation between radio and X-ray luminosity. Indeed, the present data can be described by a linear fit $Log_{10}(L_X)=\alpha\cdot Log_{10}(P_{1.4 \rm GHz})+\beta$, where $\alpha$ and $\beta$ can only be determined with very large uncertainties: $\alpha=0.5^{+1.5}_{-0.5}$, $\beta=-10^{+10}_{-36}$. On top of it, such an already loose correlation is entirely lost if we discard the one point at the highest radio and X-ray luminosities. This implies that, {\it despite originating from the very same source, the signals produced in the two bands result only very weakly correlated one to the other, and this also happens in the case of those AGN associated to galaxies of large, $M_*\ge 10^{11} M_\odot$, stellar masses}.

\section{conclusions}

We have presented a comparative analysis of the properties of  $z=[0-5]$ AGN emitting at radio and X-ray wavelengths and of their hosts. 
 The study was performed on a total of 907 X-ray AGN and 100 radio AGN drawn from the CDFS and UDS fields and endowed with information from new and ancillary data available to the VANDELS collaboration.
 Radio AGN were identified by using the method introduced by Magliocchetti et al. (2014) based on radio luminosity alone. X-ray AGN in the UDS were instead classified by using the simple criterion for the X-ray luminosity $L_X\ge 10^{42.5}$ erg sec$^{-1}$, while for X-ray AGN in the CDFS we adopted the classification presented in Luo et al. (2017).\\
  Investigations of the two X-ray AGN samples show that the one obtained on the CDFS is complete for luminosities, $L_X\simgt 10^{42}$ erg sec$^{-1}$, while that on the UDS for $L_X\simgt 10^{43}$ erg sec$^{-1}$. In spite of the different depths of the radio surveys performed on the two fields, we instead find that the radio-AGN samples originating from the CDFS and UDS have similar properties in terms of size and completeness levels, at least up to $z\sim 3$.
 
As a first result, we find that the evolutionary properties of X-ray selected and radio-selected AGN are rather different, as the distribution of this former population tends to remain flat up to $z\sim 2$, while that of radio-AGN peaks at much lower, $z\sim 1$, redshifts and starts declining already in the relatively low-redshift universe.
In spite of this finding, our data also indicates that the chances for simultaneous radio and X-ray emission from the same AGN are independent of look-back time or, in other words, that {\it given two underlying distributions of AGN active at radio or  X-ray wavelengths, the probability that an AGN will be simultaneously active in both bands is the same at all cosmological times.}

Our analysis also shows that the mass of the host galaxy is a fundamental quantity which determines the level of activity of the sources at the various wavelengths.
Indeed,  large stellar masses are found to be connected with radio activity of AGN origin, as virtually all radio-emitting AGN in our sample are hosted within galaxies of  $M_*> 10^{10} M_\odot$. Large stellar masses also seem to favour AGN activity at X-ray wavelengths, even though X-ray AGN present a more spread out mass distribution which features a sharper decrement for  $M_*\simgt 10^{11} M_\odot$ and a non-negligible tail  (37 AGN out of 55 objects in the CDFS) at low,  $M_*\simlt 10^{9} M_\odot$, masses.

Stellar mass also seems to play an important role for what concerns simultaneous radio and X-ray emission. 
Indeed, we find that the percentage of AGN which simultaneously emit at both wavelengths increases by a factor $\sim 9$, from around 1.5\% of the X-ray-selected  AGN population inhabiting galaxies of masses $\le 10^{11} M_\odot$ to $\sim 13$\% amongst more massive galaxies. In the case of radio-selected AGN, such a percentage moves from $\sim 15$\%  to $\sim 45$\% within galaxies with $M_*\ge 10^{11} M_\odot$. We stress that these are average values, and in the case of deeper X-ray observations such as those carried out on the CDFS, the percentages of AGN with simultaneous radio and X-ray emission within massive galaxies respectively rise to $\sim 17$\% of the parent X-ray AGN population and to $\sim 80$\% of the parent radio-AGN population. 

No other quantity investigated in our work seems to be connected to an enhanced probability of having simultaneous radio and X-ray emission from an AGN: neither cosmic epoch, nor radio luminosity, X-ray luminosity, Eddington ratio or even star-formation rate of the host galaxies. Furthermore, in agreement with the Merloni, Heinz \& Di Matteo (2003) results for their high-luminosity sources, there only seems to be a very loose relation between radio and X-ray luminosity in all those AGN which emit at both frequencies. In other words, {\it despite the fact that the two signals originate from the very same source, they appear to be only loosely correlated one to the other}. This happens at all galaxy mass scales, also within the more massive objects probed by our analysis.

In the light of the above results, it is not however clear whether large galaxy masses somehow favour simultaneous radio+X-ray AGN activity  
or whether radio emission is at some level present in all X-ray emitting AGN and what varies is instead the lifetime of the radio phase, resulting longer in more massive galaxies. We will investigate this issue in a forthcoming paper.\\

\noindent {\bf Acknowledgements}\\
We warmly thank the anonymous referee for constructive comments and Marta Volonteri for useful discussions which helped shaping up the paper.

\end{document}